# Dimensional metrology and positioning operations: basics for a spatial layout analysis of measurement systems

*A. Lestrade*
Synchrotron SOLEIL, Saint-Aubin, France

**Abstract**
Dimensional metrology and positioning operations are used in many fields of particle accelerator projects. This lecture gives the basic tools to designers in the field of measure by analysing the spatial layout of measurement systems since it is central to dimensional metrology as well as positioning operations. In a second part, a case study dedicated to a synchrotron storage ring is proposed from the detection of the magnetic centre of quadrupoles to the orbit definition of the ring.

## 1 Introduction

The traditional approach in Dimensional Metrology (DM) consists in considering the sensors and their application fields as the central point. We propose to study the geometrical structure or 'architecture' of any measurement system, random errors being a consequence of the methodology.

Dimensional metrology includes the techniques and instrumentation to measure both the dimension of an object and the relative position of several objects to each other. The latter is usually called *positioning* or *alignment*.

Dimensional metrology tools are split into two main categories: the sensors that deliver a measure of physical dimensions and the mechanical tools that deliver positions (centring system).

A third component has to be taken into account: time dependence of the measures coming from sensors but also from mechanical units. It is usual to consider measurement systems (whatever the techniques or the methods) as evolving in a pure 3D space. But, if ultimate precisions have to be reached, the system cannot be studied from a steady state point of view: any structure is subject to tiny shape modification, stress or displacement (e.g., thermal dependence). In other words, metrology depends on time.

Finally, space is obviously to be considered. The three-dimensional geometry (affine and vector spaces) is central. From this point of view, we could define the topic as the spatial layout (or topology) analysis of any measurement system. As an introduction, let us consider a metrology loop similar to a tolerance stack-up of a complex mechanical assembly: such metrology loops have necessarily a three-dimensional aspect imposed by the relative position of parts to each other.

Spatial analysis can be used for fields other than dimensional metrology: the design of a magnetic bench, i.e., the choice of the technology (coil, Hall probe, etc.) needs the metrology dedicated to magnetism but the complete bench design will also includes a spatial analysis of the whole set-up.

The relationship between the four components of dimensional metrology can be summarized through the schematic in Fig. 1.

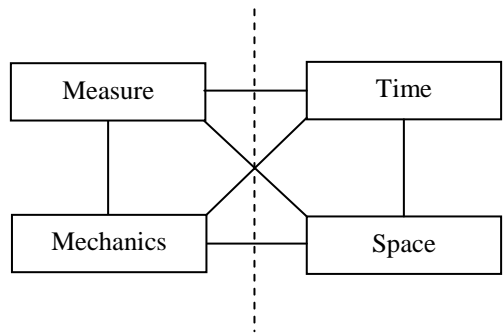

**Fig. 1:** The components of dimensional metrology

The examples of the lecture come deliberately from various fields and situations, the goal being to underline their common features and not their differences.

## 2 The sensor

Let us define the sensor in the framework of this course: the 'sensor' is the whole data acquisition chain from the physical detection to the output value, usually in a digital format (Fig. 2). However, any sensor includes a material (mechanical) part, even if it is minimal.

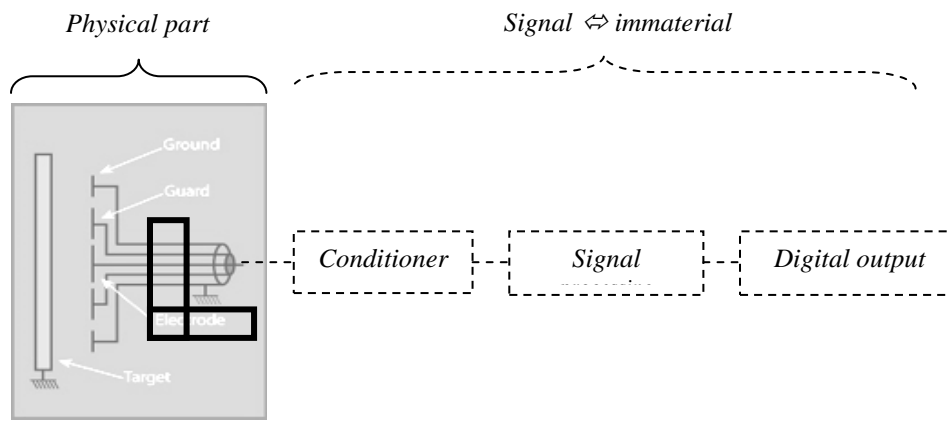

**Fig. 2:** Capacitive sensor for non-contact measurements with its data acquisition

### 2.1 Time-dependent behaviour of the sensor

The behaviour of the sensors regarding their time dependence will be linear, as a zero-order for acceleration: $y(t) = Kx(t)$, $x(t)$ and $y(t)$, being the physical detection and the digital output, respectively. The response time of the measurement device is instantaneous.

### 2.2 Categories of errors in the domain of the measure

The errors can be distinguished according to their origin:

i) Random errors: in the case of the sensors, they correspond to microscopic effects coming from the devices of the measurement chain (detection, signal processing). Their magnitude is usually small, e.g., electronic noise of the micro-component of detection.

ii) Bias errors (also called systematic errors): an offset, i.e., a bias can exist in the result of a measurement procedure. As an example, the zero value of a sensor is rarely well known. A lack of linearity can also affect the sensor. In the field of mechanics, bias errors include a misalignment between components or a shape defect of an object. The bias errors do not depend on time and their magnitude can be important.

iii) Errors depending on external sources: they can change the environment conditions of a measuring or mechanical system. Typical examples are the temperature variations during measurement or the use of mechanical assembly, the ground settlement of a long structure such as a particle accelerator. These two examples show that the range of such external influences is extremely variable. Either it is possible to have a model of their influence and one can subtract it from the result coming from the system, or they are treated as random errors.

## 2.3  Statistical model

A 'measurement' is symbolized by a pair of figures: the measure itself and the corresponding estimation of its uncertainty due to the random errors. Here are recalled the definition of the main statistical terms: average and standard deviation of a set of measurements:

$$M = \frac{1}{n}\sum_{i=1}^{n} m_i . \tag{1}$$

$$\sigma = \frac{1}{n-1}\sum_{i=1}^{n}(m_i - M)^2 . \tag{2}$$

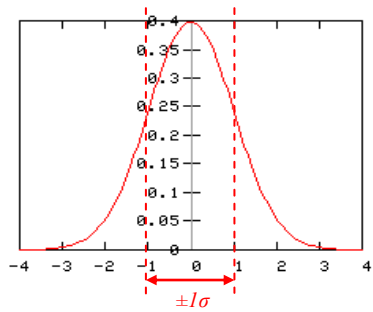

$\pm 1\sigma$

**Fig. 3:** Normal distribution of random errors

We shall restrict the study of random errors to these first-order statistical moments. Here $\sigma$ is used as an estimator of the accuracy of a measurement (Fig. 3). In the case of poor redundancy, the metrologist has to estimate the accuracy according to his own experience or from the supplier. The tolerance is defined by $T = \pm 2\sigma$. Let us assume that the measurement system is affected by a set of random errors. Let us also assume that these errors are fully independent of each other. Then, the law of random errors combination is as follows:

$$\sigma_{tot} = \sqrt{\sum_{i=1}^{n} \sigma_i^2} . \tag{3}$$

The hypothesis of independence of the random errors is often very reliable in the field of DM.

# 3 Mechanical aspects of dimensional metrology

As mentioned in the introduction, mechanics is a full component of DM as a positioning system. Even sensors include a minimum part of mechanics: the core of an electrode, its supporting part, etc. More generally, we have to consider the necessary physical part which materializes the function of a component to be positioned: the yokes, centring, and supports of a magnet for example, are dedicated to the magnetic function necessary to control the beam orbit of an accelerator. Nevertheless, it is of major importance to keep in mind that only the function of a component has to be positioned even if it is usually immaterialized, as a magnetic axis of a magnet: we do not align its yokes, even less its support.

## 3.1 Categories of errors in the domain of mechanics

### 3.1.1 Random errors

The mechanical units may be subject to uncertainty of their dimension and to external source solicitations: an assembly in a hyperstatic situation can be bent. The spatial set-up of a mechanical assembly appears at many different levels. As examples, first, the direction of a shaft depends on the clearance of the bore hole (Fig. 4). Let assume $\sigma = 10\ \mu m$ as the clearance of the bore hole, then its influence on the shaft direction is $\sigma_a = \sigma/l$ and the Z uncertainty at the point A is $\sigma_z = L.\sigma_a$.

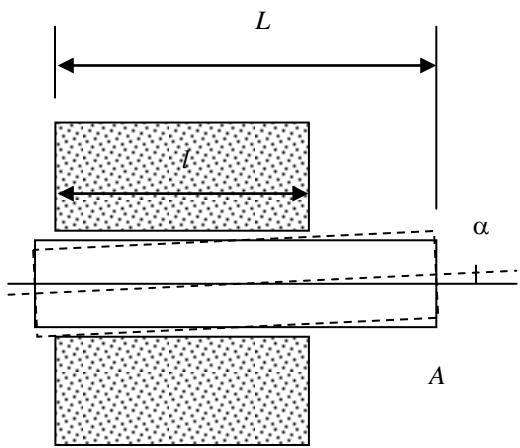
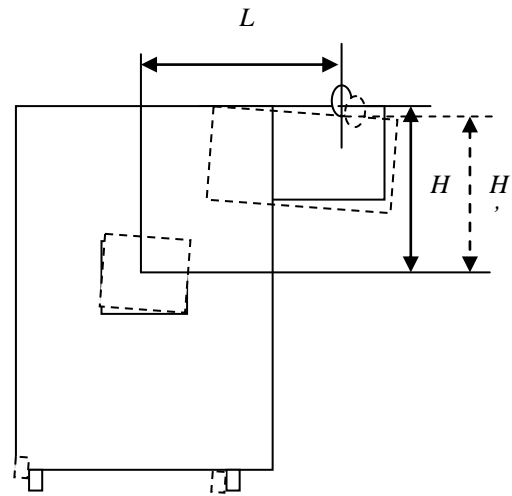

**Fig. 4:** Clearance of set shaft-bore     **Fig. 5:** Dependence on the lever arm

The height $H$ between the two points of the mechanical assembly in Fig. 5 depends on the horizontal lever arm $L$. If there is a probability $\sigma_\theta$ of parasitic rotation, then $\sigma H \approx L.\sigma_\theta$. The accuracy of machining is equivalent to random errors in the field of measurements. For that reason the tolerance stack-up of an assembly is similar to the law of error combination of measurements. However, the tolerance is preferred to the standard deviation in mechanics for practical reasons.

### 3.1.2 Bias errors

If a mechanical assembly is measured after having being machined, then the difference with respect to the nominal dimension called 'offset' can be used as a bias error regarding the whole assembly (Fig. 6).

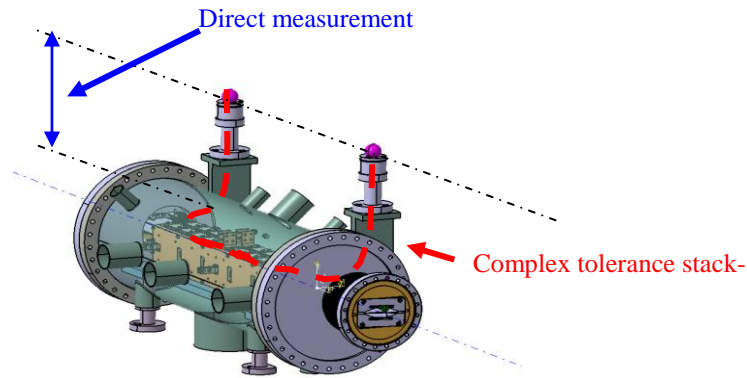

**Fig. 6:** Tolerance stack-up and direct measurement

### 3.2 Analogy between measure and mechanics

There is a formal analogy between measure and mechanics in DM. We can thus mix both sensors and mechanical units then apply the law of random errors combination and finally mix the two modes in a metrology loop. The concept of redundancy of the measures may be extended to the purely mechanical parties; there is hyperstatism in the function of positioning. Nevertheless, fully mechanical hyperstatic positioning is difficult to manage quantitatively because the mathematical modelling of 'physical' parts is not sufficient in comparison with that of the measures which are 'immaterial'. It only brings a good stability to the assembly. Conversely, a network of measures between points of a structure is, in most cases, very redundant because the theory of probability random errors applies perfectly. One can notice that applying the least-squares principle ($\sum v_i^2 \min$) in the field of measures corresponds to a minimum of energy of a mechanical system at equilibrium.

A polygon which has been measured in angle and distance is equivalent to a mechanical assembly similar to a wheel. If the measurement of the angles is weak, the wheel tends to be deformed. Strengthening the network by measures of the distance of diagonals improves its sensitivity to errors. The assembly is now more rigid (Fig. 7).

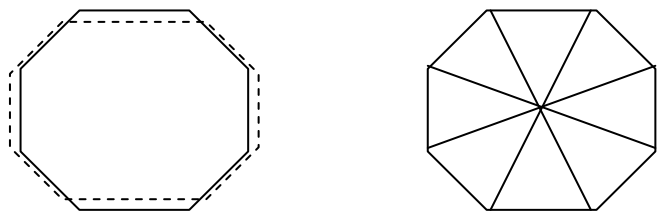

**Fig. 7:** Analogy between measure and mechanics

The analogy measure–mechanics has its own limit because of the fundamental causal relationship between both: mechanical units are made of parts which are somehow measured. After an ultimate and exhaustive analysis, one finds that the realization of the mechanical part is only a materialization of a measure by its machining step via a complex chain of actions. Similarly, carrying out the position of a mechanical set is done from preliminary measurements. However, in these two cases, the measure by which they are associated is necessarily more precise since there is always the causal relationship. Finally, after the manufacturing step, machining or adjustment at best, a measure is carried out to know the true situation (Fig. 8).

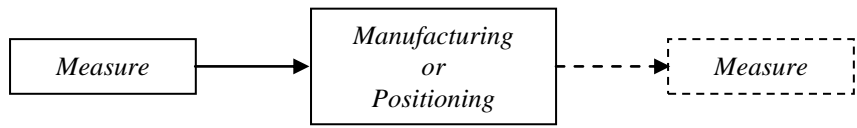

**Fig. 8:** Relationship between measures and mechanics

## 3.3 Purely mechanical aspects

The mechanical design can have a direct impact on the aspects of positioning at different stages (machining, assembly, and adjustment): the links between the sub-assemblies strongly involved in the positioning; clearance between centring systems; rolling elements not repeatable in position, etc.

i) Hyperstatism of mechanical assemblies can have a negative influence on the positioning in terms of position repeatability. Some static mechanical mounts require a hyperstatism in the functional configuration (to limit the vibration) and an isostatism for setting operation in position. Changing from one to the other cannot necessarily ensure good repeatability of position.

ii) The weight of the equipment itself must sometimes be taken into account for the machining of a large surface (Fig. 9): a beam support for magnets, whose upper face is intended to be perfectly flat, will be presented to the milling machine with its supporting points as laid down in operation; the sag due to its own weight is thus eliminated by the machining.

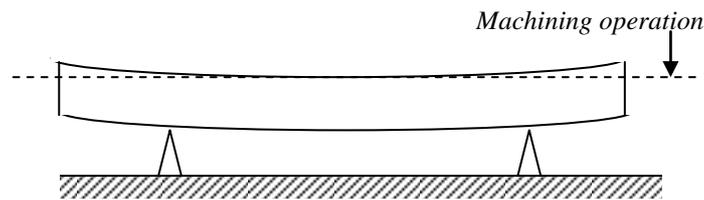

**Fig. 9:** Elimination of own weight of a girder

iii) The settings of a mechanical assembly must be made in the configuration of equilibrium closest to that in operation. It is necessary to mount all the parts, especially if they are heavy or off-set.

iv) Some parts of an accelerator must be baked out at high temperature to obtain a good quality of vacuum in the chambers. The temperature can be higher than 200°C, affecting the relaxation of the heated parts. Consequently, it is better to avoid any accurate measurement before bake-out.

## 4 Stability time constant

### 4.1 Definition

The use of a dimensional measuring instrument must take into account the stability of the set including both the instrumentation and the object to be measured or positioned. In other words, the stability analysis of the set is mandatory if ultimate accuracy is required: the stability should be at least of the same dimensional scale as the instrument precision. It is a well-known issue for metrologists, but we propose a formalization by using the concept of Stability Time Constant (STC).

The STC is defined by the acceptable duration $\delta t$ during which we do not want less than a parasitic displacement quantity $\delta d$ (Fig. 10):

$$STC = (\delta d, \delta t) \qquad (4)$$

with $\delta d << \delta m$ during the duration $\delta t$, $\delta m$ being the measurement accuracy.

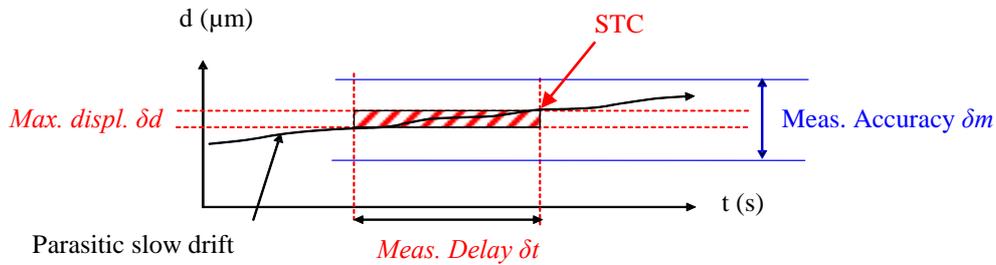

**Fig. 10:** Stability time constant

This concept can be applied whatever the origin of the disturbance of the system: mechanical, electronic, etc. The main interest of this concept is to keep in mind stability analysis at any step of the design of a measurement procedure. The STC can be defined for the six differential Degrees Of Freedom (DOF) between instrument and object.

### 4.2 Angular measurements with a theodolite

A theodolite measures angles defining the difference between two directions with, say, $3.10^{-4}$ deg accuracy. It includes a graduated circle of about ø 70 mm as an angular encoder. The duration for measurements could be $\delta t = 30$ min when there are many directions to be measured (Fig.11).

Then $STC_\theta = (3.10^{-4}$ deg; 30 min$)$.

Here $3.10^{-4}$ deg represents only 0.2 µm displacement of the graduations on the circumference of the angular sensor around the vertical axis. It shows the high level of stability required for such angular measurements.

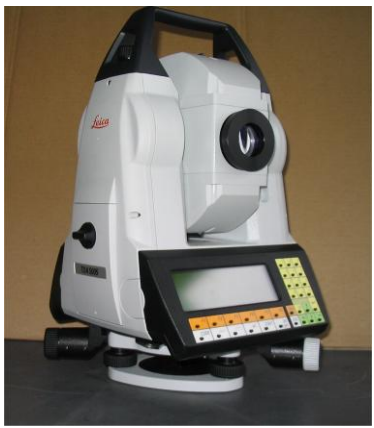 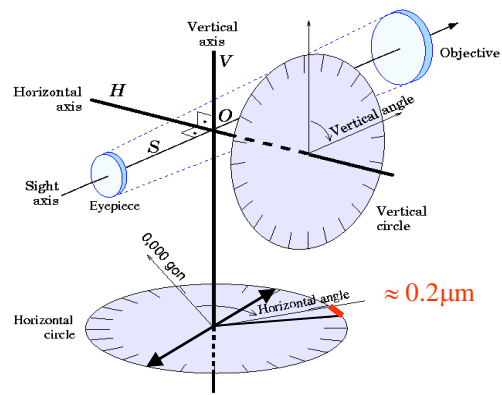

**Fig. 11:** STC of an angular theodolite measurement: 0.2 $\mu$m stability is necessary

### 4.3 Slope measurement with an inclinometer

An inclinometer measures tight slopes with respect to the horizontal plane. The measurement itself is directly linked to gravity by means of its internal sensors. Common sensors can produce measures within 10 μrad accuracy. The inner electronics needs to be stable for good results and the mechanical assembly whose physical quantity is around millimetres may vary slowly due to the thermalization of the instrument (Fig.12). Then a slow drift is always present for such instruments whose effect is equivalent to an offset on the measure origin (zero reading).

We will demonstrate in Section 7 that the half-difference of the two sensor outputs, $m_1$ in a position of the inclinometer and $m_2$ after having rotated it by 180°, removes any offset of the measurements [(Eq. 8)]: $\alpha$ is measured without offset in a minute (Fig. 12). If the measurement is carried out during the ramp of the thermal dependence curve of the inclinometer, there is no issue if the reversal is done immediately, unlike when the delay is important. Then $STC_{\theta z}$ = (10 $\mu$rad; 1 min). This is far easier to reach than the stability conditions in the example of the theodolite (Section 4.2).

In certain cases, the inclinometer can be installed for long-term measurements of a mechanical structure on which it is fixed without any capability of rotation around its vertical axis. The STC is then (10 $\mu$rad; ∞), the 'infinity' symbol meaning a long duration. It is clear that these conditions are very difficult to reach for accurate measurements.

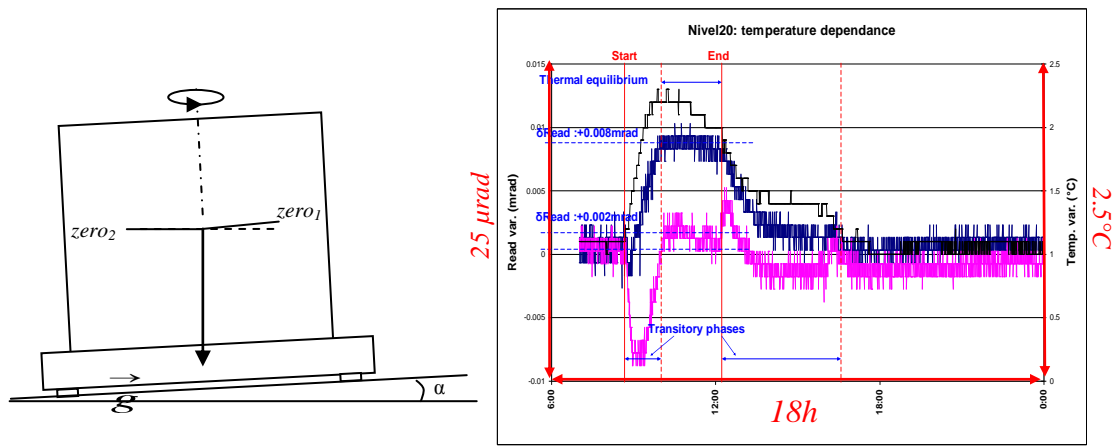

**Fig. 12:** Inclinometer rotation and thermal dependency of NIV20

The totality of instrumentation is concerned, whatever the origin of the instability (electronics, supporting mechanics, etc.). That is why every user has to regularly ($\delta T$) calibrate his instruments to keep its accuracy ($Acc$). The corresponding $STC = (Acc; \delta T)$.

### 4.4 The control of the stability issues

Many methods allow the STC to be reduced. Here are the main ones to be applied as soon as possible:

– short duration for the measurements,

– minimization of the metrology loop (Section 6),

– reversal of the instrument,

– regular calibration.

### 4.5 Choice of a referential for STC analysis

Stability analysis with STC requires a referential to define the six DOF of the system to be analysed. It is usually given by the relative position between the different parts of the unit: the distance between the object and the instrument, their relative rotation, etc. Thus in the examples of Sections 6.2 and 6.3 (translation stage), the best referential is the one defined by the direct trihedral whose abscissa axis is confused with the translation of the stage. In the example of the theodolite aiming at several targets, since the STC involves many points together, a general referential is used and described as an 'absolute'. Absolute usually means that it is linked to the Earth but the main feature of a referential is to allow the right description of the DOF of the system.

Note that the referential for calculations, if required, can be different from the one for the STC analysis. There is a particular case which involves a true absolute measurement: gravity is used for inclinometry and altimetry measurements. The metrologist has to pay attention to what can define his referential: keep it absolute or leave it as relative by differential measurement.

## 5 Geometrical frame of dimensional metrology

### 5.1 Angle–length duality

The physical space where we live is mathematically modelled by an affine space with three dimensions. The study of the mathematical operators existing in an affine space leads one to be interested in the only class of interest for DM: the displacements, i.e., rotations and translations. There is only one physical dimensional quantity in an affine space, the length, 'quantity **with** a dimension and with a unit', the metre. That is not the case of the angle, 'quantity **without** dimension and with a unit', the radian. A basic geometrical figure, the triangle, resumes clearly these properties. Three quantities are enough to define a triangle, except for the case where the three angles are known: its shape can be defined and not its dimensions.

B. Schatz wrote about the metrology of angles [1]: "*The abstract nature of the angle unit assumes an absolute accuracy. In addition, it stays available for everybody in the laboratories as in the industry. But difficulties arise when it is necessary to materialize this standard by an instrument. It is at this level of considerations that appears a certain duality, both in the means and in the methods of control, between:*

  – *the means and methods attached directly to the abstract nature of the angle unit,*

  – *the means and methods attached more directly to the unit of length and indirectly the angle unit by its trigonometric lines, those angles are thus generated by variation of length. Due to this duality, the metrology of angles is deeply linked to the metrology of the lengths.*"

Methods vary according to the environment and accuracy: one can use angles to determine lengths and one can determine angles with the use of lengths (Fig. 13). As an example, alignment, i.e., the positioning of the components of a (quasi-)linear structure such as an accelerator, can be reached either with angle measurements of a theodolite placed in the continuation of structure corresponding to tangential measurement, or with distance measurements of an EDM on the edge of the structure corresponding to radial measurements.

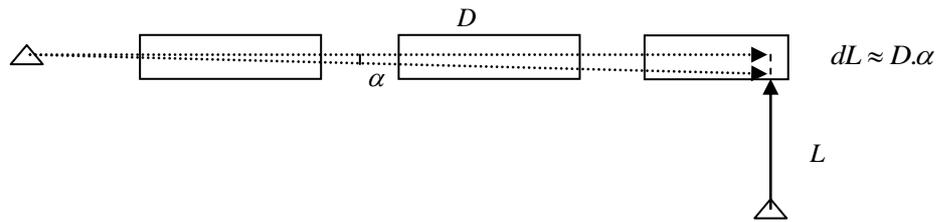

**Fig. 13:** Angle–length duality

This duality occurs even in the calculations after measurement. When an operation requires the two kinds of quantities, a least-squares calculation may be useful to the quality of the results. One can use the kind of quantities, angle or distance, in the equations describing the system. The following choice appears: equations homogeneous to angles or equations homogeneous to distances.

In some cases, the metrologist, as a user, may have to consider a vector space without length quantity, even if finally any use of angle requires that of a distance. When one wants to measure the direction of a vector, whose materialization is given by two points, then the centring of the instrument on a point and the aiming at the target of the other one, both intervene directly in the measure. Using a vector defined by two points means working into an affine space (Fig. 14).

Conversely, any measure involving autocollimation (guidance of a mirror plan, reciprocal collimation of instruments, etc.) means working in a vector space since this method is completely free of the concept of distance at the level of the user (Fig. 15); instruments and/or mirror can be positioned anywhere in the space to carry out this measure, provided that there is enough beam for the quality of measure.

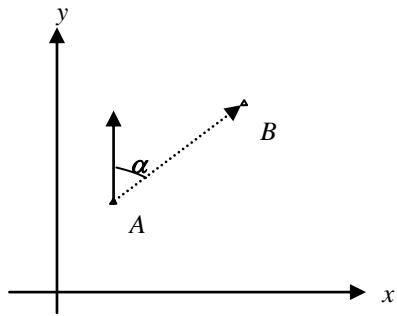
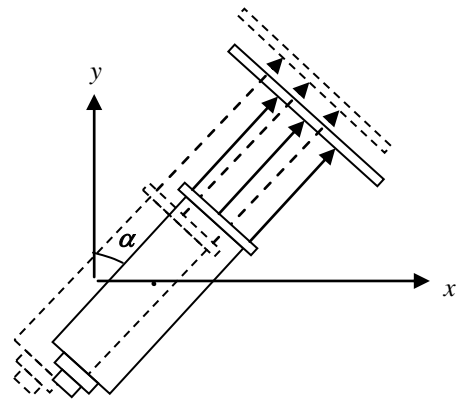

**Fig. 14:** Orientation and affine space  **Fig. 15:** Autocollimation and vector space

## 5.2 Sensitive direction

Many sensors have a unique direction of measurement. The typical example is the dial gauge: any transversal component of the part displacement cannot be detected by a dial gauge (Fig. 16). The sensitive direction of the sensor is given by its longitudinal axis. A second sensor, orthogonal to the first one, can be set for more complete information on the displacement; it is equivalent to ($dx$, $dY$) co-ordinates. In the latter case, one can calculate the displacement of the part in any other direction which can be the one of the user by the following equation:

$$R(\theta) = dX \cos(\theta) + dY \sin(\theta) .\tag{5}$$

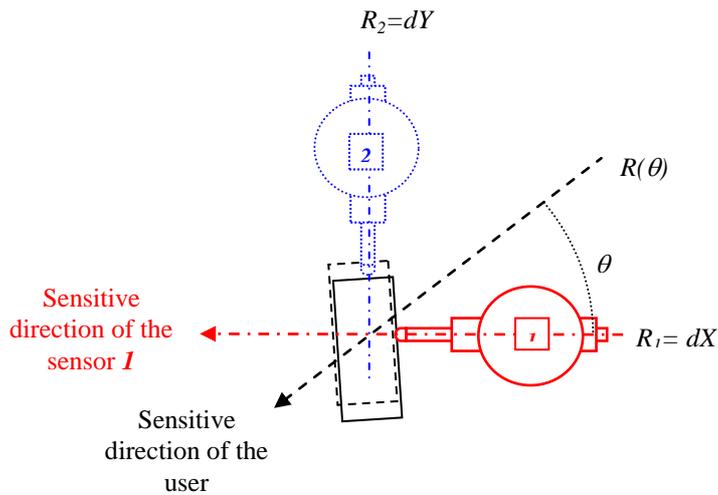

**Fig. 16:** Sensitive directions of the sensor and of the user

## 5.3 Spatial intersection

The theodolite enables 2D or 3D spatial intersection by measuring angles from two or more known observational stations to a point to be determined [(Fig 17 (a)]. That approach is very common, in DM as well as in other fields such as astronomy, physics, optics, etc. More generally, having several 'points of view' onto a problematic gives more consistent information.

Photogrammetry based on digital cameras is widely used at CERN for large physics detectors such as ATLAS or CMS [2].

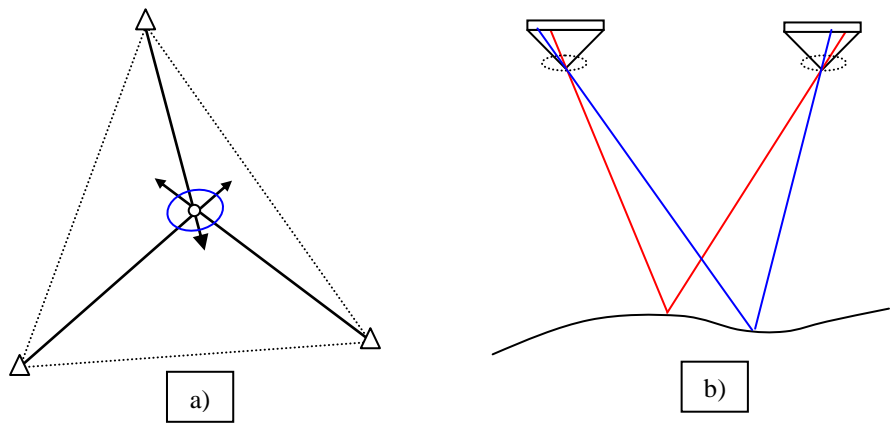

**Fig. 17:** (a) Angular intersection and error ellipse with theodolite measurements, (b) schematic of photogrammetry

In the field of surveying, spatial intersections tend to have the same accuracy in all the directions of, say, the plane in 2D geometry. One defines the error ellipse of the point to be determined in the plane as the probability to find it inside with 86% confidence (Fig. 17). This probability corresponds to $T = \pm 2\sigma$ (see Section 2.3) which is the standard deviation of a set of measurements applied to 2D. A 'good' layout in standard surveying must give an ellipse closed to a circle whose radius is as small as possible.

But what is linked to the concept of sensitive direction is the fact that sometimes one cannot be interested in an isotropic information: aligning quasi-linear structures such as an accelerator leads to intersections with angles close to zero or to $\pi$ radians (Fig.18). The small axis of the error ellipse and therefore the sensitive direction of the angle instrumentation used are then in the radial direction of the accelerator beam axis for alignment purposes.

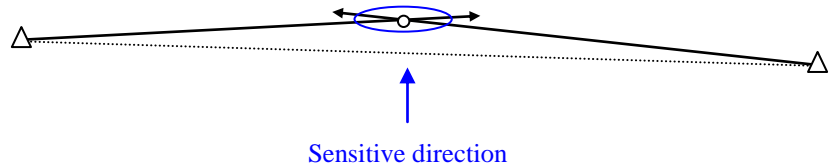

Sensitive direction

**Fig. 18:** Angular intersection and error ellipse of the point to be determined in the accelerator field

The use of a laser tracker with its accurate interferometer leads to the same principle of error ellipse and sensitive direction applied to distance measurements carried out in the normal direction of the beam axis.

## 5.4 Effective length

In the case of small angle measurement like inclinometry, the concept of 'effective length' is very useful. The mechanical design of such an instrument requires a stack of interfaces from the sensor to its stands in contact with the part to be measured. The user, who has to estimate the accuracy or the STC of a system, must keep in mind the fact that whatever the length of the part to be measured, the effective length of the physical detection is at most equal to the size of the sensor body. It is actually usually far smaller for inclinometers. Monitoring a structure with a fixed inclinometer at the microradian scale needs the corresponding stability for a very long duration: $STC = (\mu rad, \infty)$. That stability constraint, being applied for the whole stack of mechanical interfaces, is obviously true for the smallest one, which is usually the detector itself. Such support is about 10 mm: that is the effective length of physical detection. Reaching 1 $\mu$rad with a 10 mm lever arm is equivalent to limiting any parasitic vertical displacement of the detector bigger than 10 nm. Conversely, using HLS (non-contact displacement sensors) on a free surface of water for inclinometry (Fig. 19) can achieve easily 0.1 $\mu$rad because the effective length is the distance between the sensors, i.e., some hundred metres.

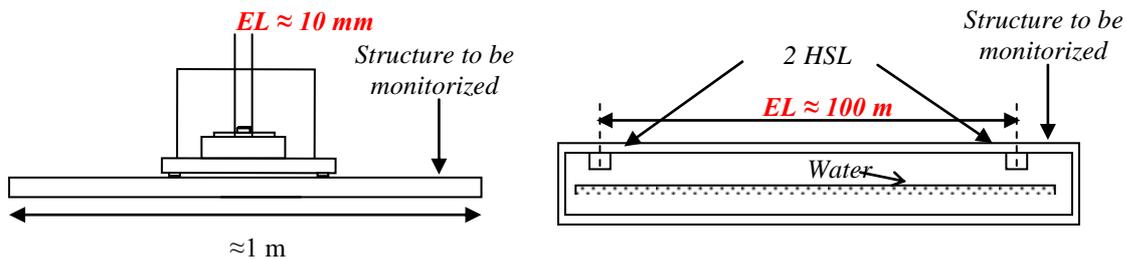

**Fig. 19:** Effective Length (EL) in inclinometry

Another example is the case of lamination machining for a dipole magnet (Fig. 20). The shape or size tolerance is typically ±0.02 mm required by the designers for any point of the lamination perimeter. A usual mistake is to believe that the accuracy of the mechanical tilt (rotation around the beam axis) of the magnet is $0.02/L = 0.025$ mrad because the lateral range of the yoke is its width $L = 786$ mm. The effective length for the electron beam is actually the width of the pole $p = 128.6$ mm which is smaller. Then, the tilt accuracy due to machining is $0.02/p = 0.156$ mrad (per pole).

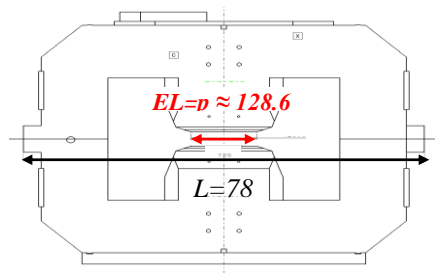

**Fig. 20:** Effective Length (EL) of a dipole tilt

## 5.5 Sine and cosine errors

The sine error usually called Abbe error, is the one committed in a length measurement whose reference axis is shifted with respect to its displacement axis or to the contact on the object to be measured (Fig. 21): any parasitic rotation θ between the instrument and user axes induces a 'sine' error on the result:

$$e_{sin} = d . \sin\theta \qquad (6)$$

where *d* is an offset between the two axes called 'Abbe length'.

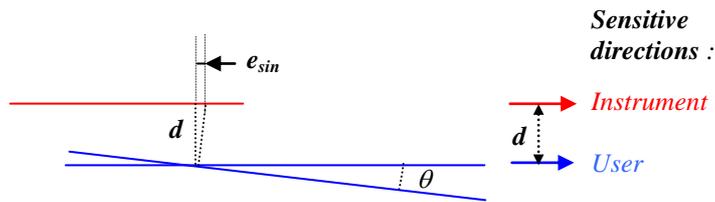

**Fig. 21:** Abbe or sine error

Ernst Abbe (1840–1905) wrote: *'to realize a good measure, the measurement standard must be placed in the same line as the dimension to be checked'*. In Fig. 22, the interferometer used to calibrate the unit is properly set at the same level as the point to be checked ($d = 0$) while the optical ruler shows an Abbe length *d*, sensitive to any parasitic rotation *θ* of the stage with respect to the optical ruler.

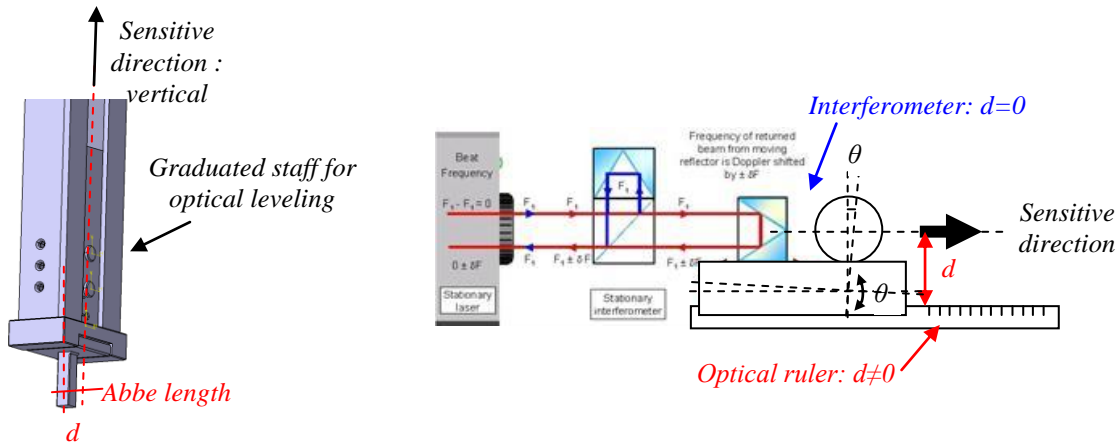

**Fig. 22:** Layouts with Abbe error

This error is often accompanied by its twin, called 'cosine error' and arising if the sensitive direction of the sensor is not parallel to the dimension to be checked (Fig. 23):

$$e_{\cos} = L.\cos\alpha \approx \frac{h^2}{2L} \tag{7}$$

where $\alpha$ is an angle between the two axes.

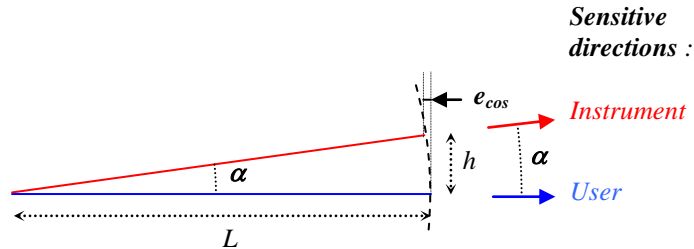

**Fig. 23:** Cosine error

### 5.5.1   Calibration of a linear displacement sensor using an interferometer

Consider the previous example of the interferometer checking a translation stage without any Abbe error. Suppose now that the Abbe length is $d = 2$ mm. That length applied to a maximum rotation $\theta = 0.10$ mm/m due to the pitch of the sensor head when it moves on the optical rule gives an Abbe error of about: $e_{max} = 0.002$ mm.

On the other hand, if the direction of the interferometer is rotated by $h = 1$ mm applied to $L = 50$ mm to be measured, then the cosine error is

$e = \dfrac{1^2}{2\times 50} = 0.010$ mm. In this example, the cosine error prevails.

### 5.5.2   Combination of sine and cosine errors

If both errors are combined, the geometric formulation corresponds to a rotation in 2D (Fig. 24) and the matrix formulation is

$$M' = \begin{vmatrix} x' \\ y' \end{vmatrix} = \begin{vmatrix} x\cos\alpha - y\sin\alpha \\ x\sin\alpha + y\cos\alpha \end{vmatrix} \approx \begin{vmatrix} -y\sin\alpha \\ x\sin\alpha \end{vmatrix}_{x\approx y} \tag{8}$$

where $(x')$ and $(y')$ are the horizontal and vertical sensitive directions respectively.

The sine errors prevail if $x \approx y$.

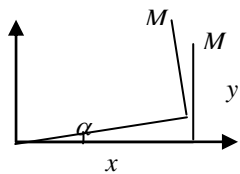

**Fig. 24:** Combination of sine and cosine errors

## 5.5.3 Fiducialization of a quadrupole magnet

Any fiducials of an accelerator component show such lever arms (*d* and *h* in Fig. 25) with respect to the magnetic axis, and therefore sine errors versus $\theta$ the rotation around the beam axis (tilt).

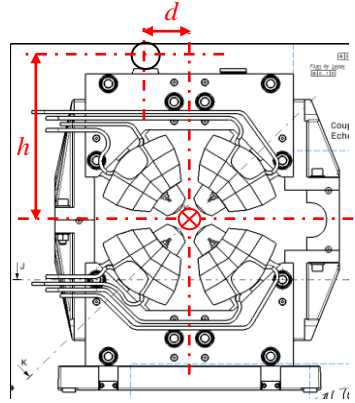

**Fig. 25:** Lever arms of quadrupole fiducials

## 5.5.4 The control of lever arms

The Abbe principle being difficult to apply, it is necessary to control the lever arms *d* and the parasitic rotations $\theta$ even in the three dimensions of space.

Three classes of solution exist according to the system of measurement to limit the sine error:

i) The sensor is in the axis of measurement (*d* = *0*).

ii) Several sensors are placed symmetrically with respect to the axis of measurement; the sine function being odd, the average of the measurements delivered by the sensors corrects the error.

iii) An angular sensor manages any parasitic rotation.

Two classes of solution exist according to the system of measurement to limit the cosine error:

i) The sensor is in the axis of measurement (*h* = 0).

ii) An angular sensor manages any parasitic rotation.

It is important to note that multiple sensors placed symmetrically with respect to the axis of measurement do not correct this error, the cosine function being even.

# 6 Metrology loop

Any system dedicated to positioning or requiring a positioning operation for its good working consists of a succession of mechanical parts and/or of sensors. This succession of elements is called 'metrology loop': it can be seen as the support of position information transmission [3]. Several kinds of metrological chain are used according to whether it is dedicated to the position by mechanical means or is a set of measures spatially distributed. The first one is sometimes called *mechanical, assembly or actuator loop* according to the case; the second one is sometimes called *measuring loop*. Finally, if we must analyse a system of positioning from the point of view of stability in order to integrate the effect of a parasitic slow drift, then it is necessary to take a *stability loop* into consideration. However, that vocabulary must remain flexible; any rigorous application of these names fails with the following example: a measuring loop necessarily requires mechanical parts.

The *metrology loop* is the best structure for a calculation of error budgets, bias errors, and STC analysis. By using the word '*structure*' we can feel the relationship with a spatial layout.

### 6.1 Metrology loops dissociation

The mechanical units which need both kinematics and measurement as a set point must be designed with fully independent actuators and measuring loops. That is the only condition so that the measuring system can detect any true movement coming from actuators. The modern Co-ordinate Measuring Machines (CMM) are now designed with the condition that the two loops are fully disconnected from each other. The term '*dissociated metrology*' is used in that case [3], [4]. Applying the dissociation principle brings some interesting aspects in other cases: if the precision of a unit dedicated to positioning cannot be reached by mechanical means, then it is sometimes possible to replace it by a measurement.

### 6.2 Translation stage with coaxial micrometer

The reading of the micrometer is used as set point information for the translation of the stage to the required position (Fig. 26). The actuator and measuring loops are coaxial, that is they correct to eliminate Abbe error, but they are not completely independent: the backlash of the displacement screw of the stage does not appear in the measuring loop that induces a limit to the positioning accuracy.

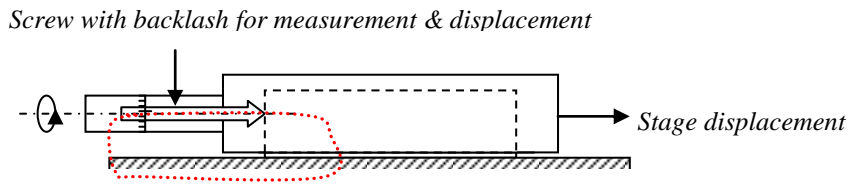

**Fig. 26:** Translation stage with coaxial micrometer

### 6.3 Translation stage with independent sensors

Both loops are clearly independent in Fig. 27. The sensor can detect the screw backlash delivering thus an accurate set point for the displacement. However, if the backlash induces rotations, then one sensor only is not sufficient.

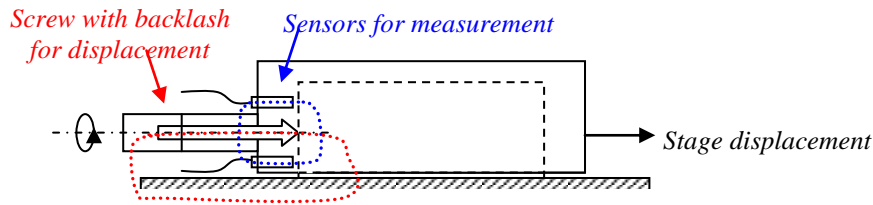

**Fig. 27:** Translation stage with independent sensors

### 6.4 Co-ordinate measuring machine and machine tool

This kind of topology called 'series' is costly in terms of accuracy on the relative position between the sensor and the part to be measured, or the tool and the part to be machined (Fig. 28). All the intermediate parts of the metrology loop must be known accurately and thus for the six DOF of any mechanical subset.

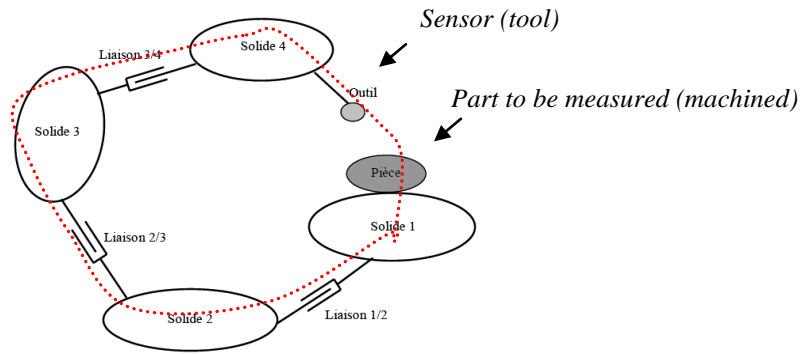

**Fig. 28:** Serial layout [3]

## 6.5 Polygonal traverse for surveying

This topology is fairly usual in the field of surveying (Fig. 29). All the distances $D_i$ and the angles $\alpha_i$ are measured from *A* to *B*. One can clearly see the analogy with the previous mechanical example. This one shows the disadvantage of such a 'series' topology: the errors pile up all along the traverse.

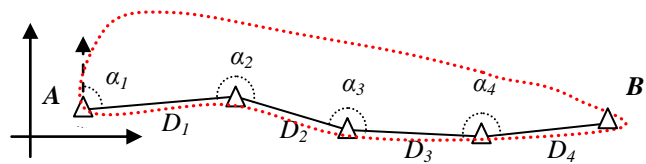

**Fig. 29:** Polygonal traverse

## 6.6 Hexapod

This kind of mechanical structure is called 'parallel' (Fig. 30). It is much less sensitive to the errors of positioning since the position of the upper stage depends on each of the six legs of the hexapod. It is important to note that this structure is not hyperstatic: it is strictly equivalent to the six DOF of a solid in space. Decoupling actuators and measuring loops is extremely effective in such topologies [3].

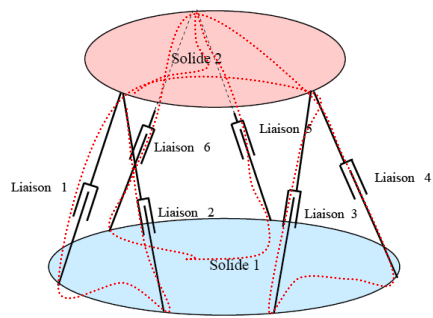

**Fig. 30:** Parallel layout [3]

## 6.7 Optical measurement

When measuring a part (say, its shape) by means of an optical instrument separate from the support of the part (Fig. 31), the metrology loop includes:

– the instrument with the target to be used to define the points on the part,

– their respective supports (tripod, target support),

– the volume of the air on the optical path of the measurement of its mechanical interface,

– the support of the part,

– the ground.

Any mechanical part and any measurement which could produce errors has to be included in the metrology loop. Both STC and random errors analysis must be analysed. The metrologist has to choose his level of detail.

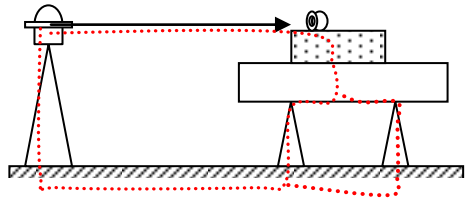

**Fig. 31:** Optical measurement separate from the part

## 6.8 Bench for magnetic measurements

A measurement bench is a facility dedicated to the calibration of series of components. In terms of positioning it locates the position of the functional part of the component with respect to a mechanical reference externally accessible for further positioning operations. The bench shown in Fig. 32 is dedicated to the detection of the magnetic axis of quadrupole magnets. Once the axis is detected by the sensor, one extends this position information to the upper yoke: this last step is called 'fiducialization'.

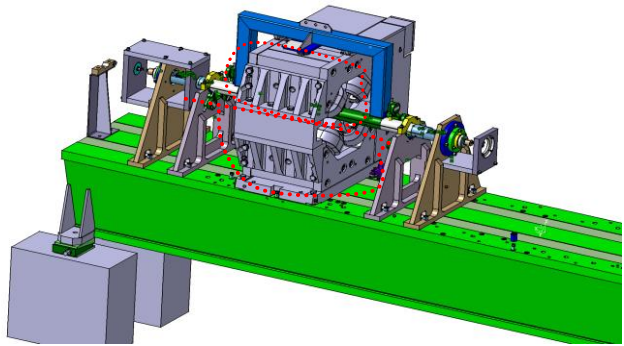

**Fig. 32:** Bench for magnetic measurement

## 6.9 Offset measurement of mechanical assembly

The relative position between the functional part and the external references (fiducials) can be assured mechanically by a drastic tolerance stack-up imposing very accurate machining and assembly (Fig. 33). It is sometimes better to release the constraints (and the cost) of mechanical machining and to measure precisely according to the chain of measurement indicated on the figure.

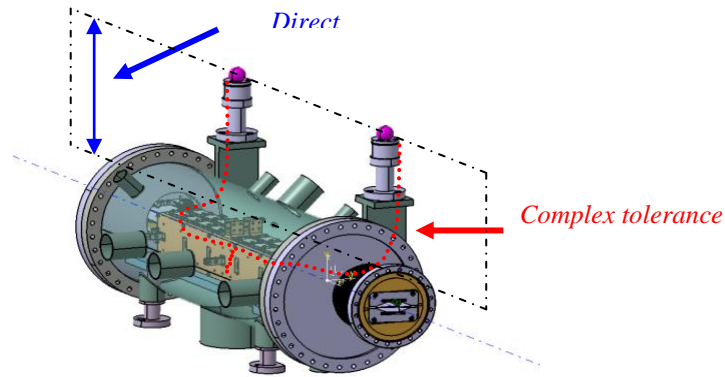

**Fig. 33:** Tolerance stack-up and direct measurement

## 6.10 Geodetic network

A network of measurements (i.e., a set of measures) to define a mechanical unit or to align them is a trivial example of a metrology loop. Since the redundancy is easier to control in the field of measurements than in mechanics, we shall use it as much as possible. The geodetic network includes all possible measurements (angles, distances, alignment, etc.) with a tacheometer between the different objects (Fig. 34).

The metrology loop is in fact more complex: the centring systems of the instruments and tools, the air where the optical beam travels and sometimes the inner sensors of instrumentation must be included.

Finally, the network may correspond to the measures between magnets installed on different girders to be aligned with respect to each other. The metrology loop includes the measured points. The complete STC analysis for the slow drifts of the whole accelerator must incorporate the girders, the pedestals of girders, and also the concrete slab on which it is installed.

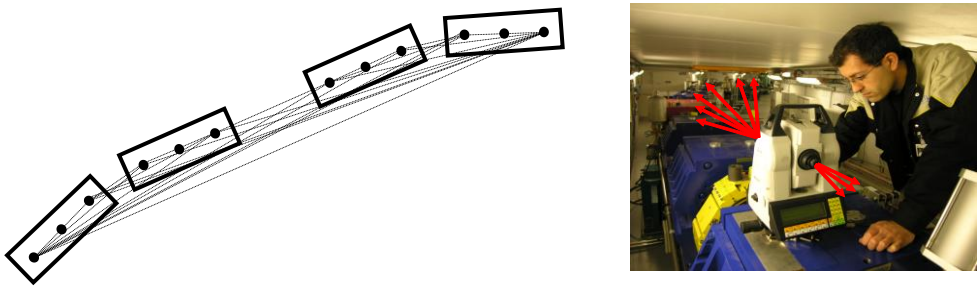

**Fig. 34:** Geodetic network and tacheometer TDA5005

## 7 Reversal layout

The error separation principle is used in any field of measurement whatever the trade, mechanics, physics, magnetism, alignment, etc. Thus that family of methods includes many different configurations but they all use the same frame called Error Separation Layout (ESL). Since this concept is of prime importance in DM, we introduce it progressively step by step.

### 7.1 The simple reversal

Any rotation of 180° makes it possible to eliminate offsets of centring systems. It is a powerful principle of bias reduction. The whole kinematical assembly that is reversed will be free of centring error. It is important to note that only the systematic part of the centring errors disappears and not the random errors, the number of which remains as many as there are reversed subsets. If the metrology loop is a series type, the errors are composed quadratically.

#### 7.1.1 *Measure of the location of a bore by means of the reversal method*

A target is used to determine the planimetric location of a theodolite triback with respect to an external reference (Fig. 35), typically for its positioning on the alignment defined by two targets of reference. Since the bore called '*A*' is not directly usable for an optical measurement, a spherical target is laid on a spacer. The theodolite aims at the target for an angle measurement.

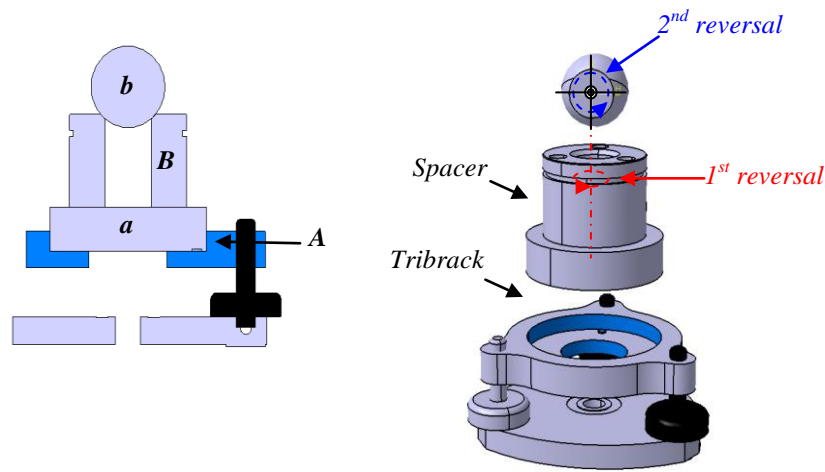

**Fig. 35:** Reversal method for centring systems

This spacer is machined at its ends as follows:
- At the bottom: a shaft called '*a*' with the same diameter as that of the bore '*A*' (with the functional clearance).
- At the top: a cone '*B*' accepting the sphere with the target '*b*'.

Two 'bias' errors can exist, related to the mechanical machining:
- The offset between the shaft axis '*a*' and the cone '*B*' of the spacer.
- The offset between target and sphere centres.

A set of angle measurements is carried out:
- $R_1$ in an unspecified configuration.
- $R_2$ after the reversal of the spacer by 180°.

The average of this set of measurements eliminates the first bias: $R = \dfrac{R_1 + R_2}{2}$. In terms of referencing, the cone *B* is linked to bore *A* of the tribrack.

By applying the same principle of reversal to the sphere with respect to its centre, the second bias is eliminated. The target is then linked to the tribrack bore; the position of the target thus becomes representative of that of the bore. The error combination of centring due to the functional clearances gives: $\sigma_{tot}^2 = \sigma_1^2 + \sigma_2^2$.

### 7.1.2 *Orientation measurement of a plane surface using autocollimation*

The orientation of a plane materialized by a machined steel surface is obtained by autocollimation on an intermediate mirror (Fig. 36). This mirror is here a tool; it belongs to the system of measurement.

The first measurement is taken in an unspecified position of the mirror provided that it is in good contact with the steel surface. A second measure is carried out after reversal by 180° around the normal direction to the plane. The average of measurements gives the good orientation of the plane:

$$R = \dfrac{R_1 + R_2}{2}.$$

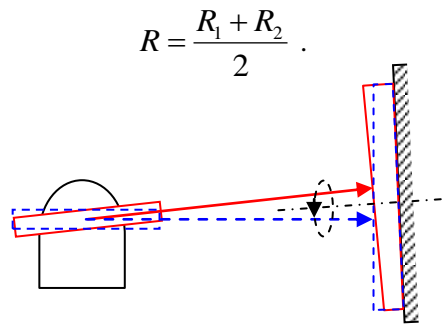

**Fig. 36:** Reversal method in autocollimation

The plane is now linked to the zero of the angular sensor of the instrument: the mirror disappears in terms of bias. If the instrument makes it possible to measure in the two angular directions (*H* and *V*), the principle of the reversal of the mirror is still valid.

### 7.1.3 *Inclination measurement of a plane using inclinometry*

One wants to measure the angle of inclination in a given direction of a plane materialized by a machined surface. An inclinometer is laid on the plan surface. The inclinometer functions with gravity which is its reference of measurement. The instrument shows a constant error (offset): the mechanical stands of the instrument define a horizontal plane for a reading different from zero.

A first reading is done, then another one after having reversed the instrument. The value of slope will be given by

$$R = \dfrac{R_1 - R_2}{2}. \qquad (9)$$

That formula corresponds to the average formula with a negative sign applied to the reading $l_2$ since the instrument and its zero are reversed and not the object to be measured. The instrumental error (offset) is given by application of the average of measurements:

$$Offset = \frac{R_1 + R_2}{2} . \qquad (10)$$

## 7.2 Multi-reversal measurement

The reversal principle can be generalized at various degrees. An important stage of this generalization is known as the multi-reversal method [4], [5].

It was developed for the dimensional measurement of parts having symmetry of order $N$. A mechanical part is entirely measured $N$ times by a coordinate CMM in the $N$ positions of the part after each rotation of $360°/N$ around its axis of symmetry (Figs. 37, 38). At each step of the part rotation, a full rotation of the CMM sensor is carried out for rounding error measurements.

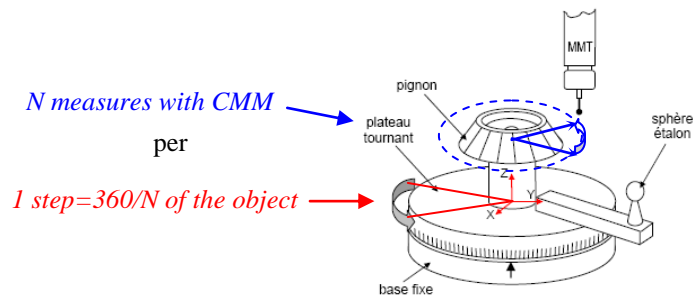

**Fig. 37:** Schematic of multi-reversal method [6]

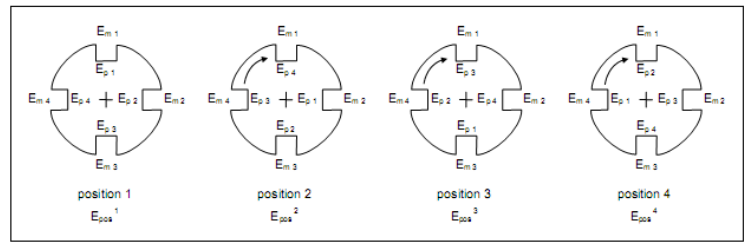

**Fig. 38:** Multi-reversal with order 4 symmetry [6]

Thus, each point of the part 'sees' successively the defects of the system of measurement and each position of measurement of the CMM 'sees' successively the defects of the part. The total error in each point is the sum of the error of the part ($E_A$) and of the measurement system ($E_M$):

$$E_t = E_A + E_M .$$

Repositioning the part in the reference frame of the CMM after a rotation induces an additional unknown common to the corresponding set of measurements ($E_P$).

$$E_t = E_A + E_M + E_P .$$

## 7.3 More layouts

The CMM can be functionally described as a whole set of *N* sensors interdependent with the same structure and measuring the parts *N* times in parallel in *N* points of measurement (Fig. 39). Each sensor shows its own error equivalent to the ones of the CMM; these errors vary with the CMM geometry of its inner mechanics, the defects due to guidance units of the measuring head. The CMM errors calculated at the *N* points of the part are a combination of them. Conversely, the study of all the error vectors after calculation can lead to a determination of guidance unit defects [6].

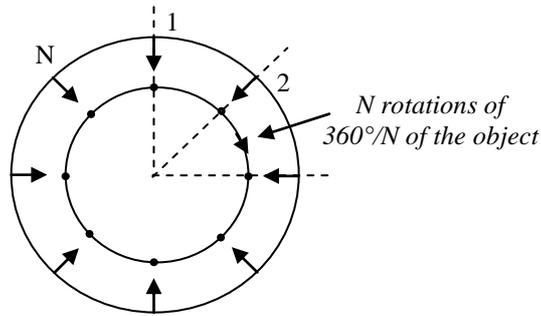

**Fig. 39:** ESL in circular configuration

The rotation of $360°/N$ is relative between part and system of measurement: any part can turn or can remain fixed, according to hardware configurations. This circular architecture can be analysed linearly if its circumference is considered (Fig. 40). The functional diagram is as follows.

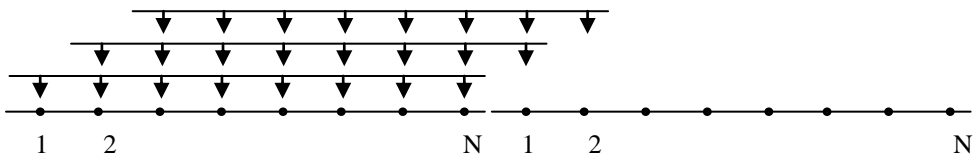

**Fig. 40:** ESL in circular configuration (circumference)

It reveals that the method can be applied to linear structures as well, without any particular symmetry, neither for the part, nor for the protocol of measurements. What is important is the ability of such an arrangement of both part and instrumentation, which allows the separation of several errors. In the rest of this lecture, we use the term 'Error Separation Layout' (ESL).

Figure 41 shows the corresponding diagram for a linear structure of *N* parallel measurements: since the symmetry does not exist any more, it is better to move the measurement structure in both directions.

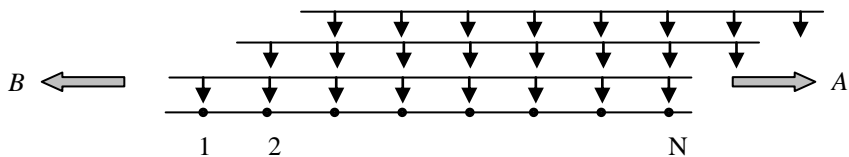

**Fig. 41:** ESL in linear configuration

The sensors can deliver physical quantities other than length. As an example, it could be the output of an angular encoder at different points of the graduated circle of a theodolite (Fig. 42). The information is no longer a radial circle variation but a tangential variation of the spacing between graduations. This variation, reported to the radius of the graduated circle is interpreted in terms of angular quantity.

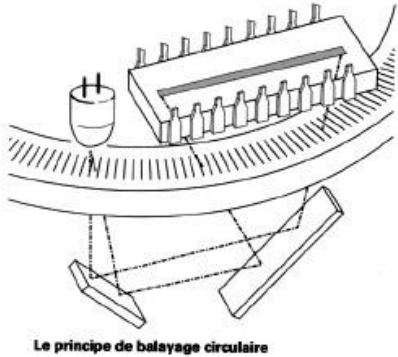

**Fig. 42:** Angular encoder of theodolite (™Leica)

Imagine that the gear wheel is now placed on a simple rotation stage equipped with an angular encoder. The roundness error measurement can be carried out by means of displacement sensors. *N* sensors for *N* measured points is expensive (but exhaustive). Using few of them or even only one is possible (Fig. 43).

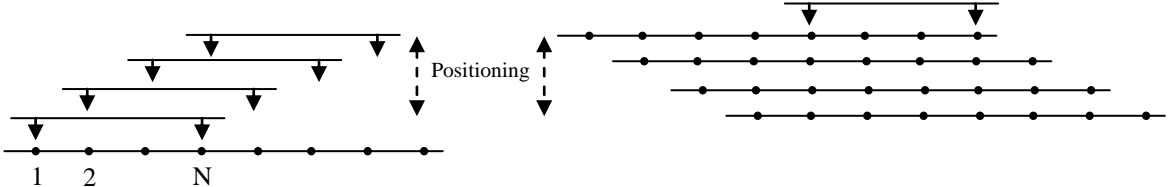

**Fig. 43:** ESL with parse layout

Note that in Fig. 43, both layouts are equivalent regarding the positioning part: the number of unknowns is the same.

# 8    Modelling of an Error Separation Layout (ESL)

Any ESL can be summarized with the following variables:

($A$, $M$, $P$, $I$).

    – the part called artefact     : $A$

    – the part corresponding to the measurement system     : $M$

    – the part or the operation for positioning     : $P$

    – the part corresponding to the instrument set-up     : $I$

Different layouts can exist. The type of the information can be radial or tangential on circular or linear supports (Fig. 44). The dimensional information can come from the probes (typically displacement sensors) or from the artefact as with the angular encoders or graduated rules.

There are typically two families of unknowns between *A*, *M* and *P* parts. In addition, the instrument set-up unknowns are either supposed to be negligible or not.

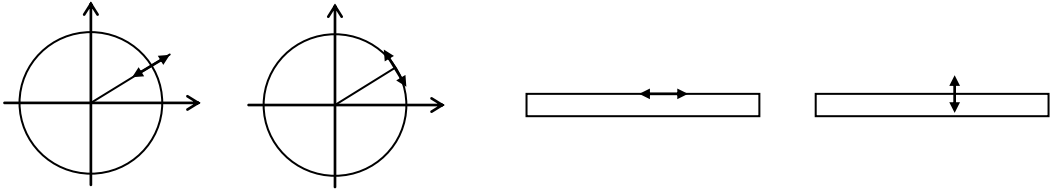

**Fig. 44:** Type of information

However, a particular case is of prime importance due to the wide range of applications and to its theoretical interest: ESL for circular measurements. The relationship between parts is shown in Fig. 45.

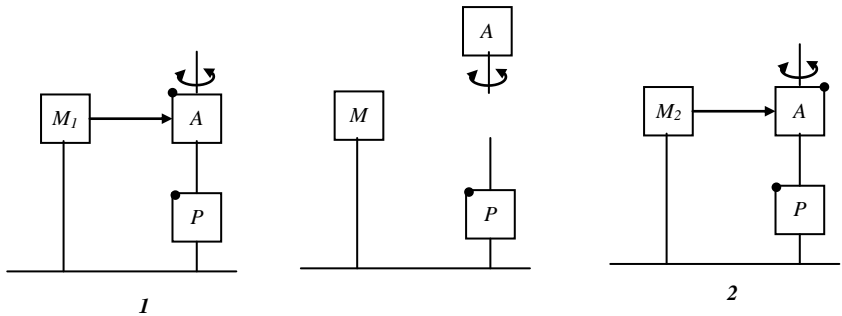

**Fig. 45:** A first attempt at ESL in circular measurements

The measuring probe is supposed to be free of any systematic errors. Then, only artefact and positioning parts errors have to be assessed. We also assume that all these errors (or variations) are repeatable whatever the number of turns the system can do. The errors are only a function of $\theta$.

The artefact is usually fixed on a spindle (or any rotation stage) whose rotation is not perfect regarding the level of accuracy to be reached on the artefact. That spindle corresponds to the positioning part of the ESL.

A unique probe for measuring the roundness errors of the artefact is set on the table. *N* points are measured with configuration no.1 of Fig. 45. Then, the artefact is dismounted and reversed on the spindle and the *N* points are measured again with configuration no. 2. Unfortunately, the reversal of the artefact on the spindle is not a true reversal with respect to the measuring probe: measurements will give partial information, in contrast to the positioning part where reversing it leads to (quasi-)complete information (see Section 8.4).

$$M_1(\theta) = A(\theta) + P(\theta) .$$

$$M_2(\theta) = A(\theta - \pi) + P(\theta) \neq -A(\theta) + P(\theta) .$$

Another 'object' appears, superimposed on the spindle parasitic movements and the shape of the artefact: the centring of the artefact on the spindle. If an eccentricity exists, it is not possible to fully reverse it. It creates a sin $\theta$ curve superimposed on the measurements in the probe output. It is included in the instrument set-up '*I*'.

## 8.1 The positive ESL model

The complete model is shown on Fig. 46. It includes the following families of objects:

- Artefact: only a variable per $\theta$, typically the radius $r(\theta)$ for radial measurements, or the angle $\alpha(\theta)$ for tangential measurements.
- Positioning: two variables per $\theta$: $X(\theta)$, $Y(\theta)$.
- Instrument set-up: $(dX, dY)$ for the eccentricity, its effect is as sin $\theta$, corresponding to the harmonic $n = 1$ in Fourier analysis.
- Measuring probes: $e(o)$, $o$ being the probe output.

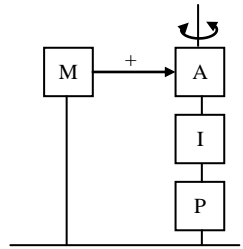

**Fig. 46:** The ESL$^+$ model

The direction of the arrow symbolizing the measurement will make sense later on (see Section 8.10). We propose to define the sign '+' as 'direct measurement' for such an arrangement. The use of angular encoders will be with the negative sense.

The nature of artefact and positioning are clearly different.

Note that these objects can be extended to linear layouts with the corresponding variables. For example, the artefact can be described by $r(l)$ for straightness errors, $L(l)$ for graduation errors (Fig. 44).

There are usually only two families of unknowns: *A* and *P* in the example of the roundness and spindle errors measurements, *A* and *M* in the case of the gear wheel on the CMM.

## 8.2 The symmetry of true reversal

Evans et al. [7] showed that the reversal is mathematically similar to symmetry, usually obtained from a physical $\pi$ radians rotation of one part of the layout to be assessed with respect to the measurement system, for example, the artefact or the positioning part. The study of the invariant allows the detection of reversal layout:

i) **Artefact**: If the probe does not move, the invariant of a circular artefact is, say, the mean circle. Changing the direction of the error curve is clearly impossible (Fig. 47, index 1). The $\pi$ radians rotation of the artefact is not a reversal because the required symmetry is not related to the centre of rotation of the artefact but to its mean circle. Another layout allowing a reversal condition is to rotate artefact and probe, then change the direction of sensitivity of the probe (Fig. 47, index 2). The invariant of the symmetry is here a point defined by the probe itself. Finally, a rotation of the artefact around a horizontal axis with the change of the sensitive direction is also possible (with the change of direction of $\theta$ rotation), as in straightness measurements (Fig. 48) of a slideway with respect to a datum straight edge.

ii) **Positioning part**: the invariant is a point in 2D, the $\pi$ radians rotation of the complete spindle (rotor + stator) is a reversal layout (Fig. 47).

iii) **Eccentricity**: as for the trajectory of the positioning part, the invariant for symmetry is a point in 2D. In any case, repeating it after dismounting the artefact at micrometre accuracy or obtaining its reversal configuration with a high level of precision are major issues. As a consequence, the eccentricity cannot be separate from the *A* and *P* parts (Fig. 49).

iv) **Measuring part**: In some cases, the parity of the linearity curve of probes can lead to symmetry.

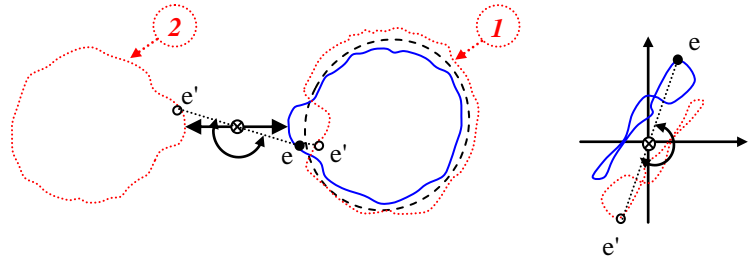

**Fig. 47:** Invariant of artefact and positioning part for reversal in circular measurements

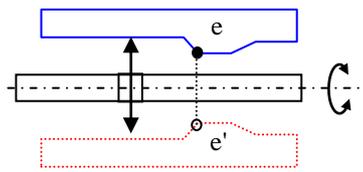

**Fig. 48:** Invariant of artefact for reversal in linear measurements

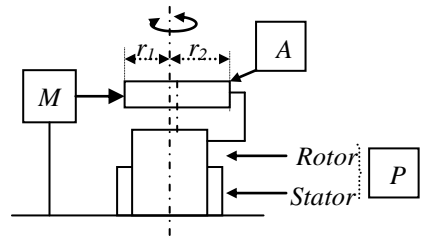

**Fig. 49:** Eccentricity of artefact/spindle

## 8.3   Extension to 2D and 3D problems: set of points remaining invariant

Considering the $\pi$ radians rotation and its invariants is a first step in a more global approach applied to 2D and 3D problems. Self-calibration methods of stages dedicated to lithography in the semiconductor industry are based on the use of a lattice of invariant points defined by its symmetries [8].

## 8.4   The Donaldson reversal

When the positioning part (i.e., the full set rotor + stator of the spindle) can be reversed with respect to the probes and to the artefact, the situation is perfect: measuring all the *N* points versus $\theta$ and measuring again after the rotation of the spindle, leads to artefact and positioning errors in the sensitive direction of the probe (Fig. 50) without any loss of information except for harmonic $n = 1$ because of the lack of knowledge of the instrument set-up:

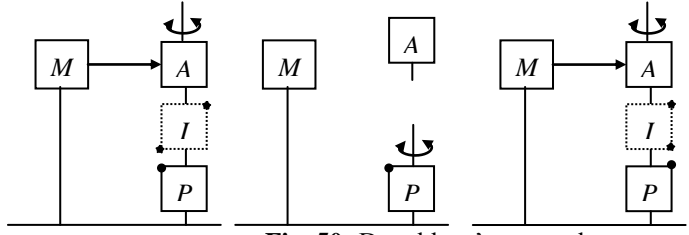
**Fig. 50:** Donaldson's reversal

$$M_1(\theta) = A(\theta) + P(\theta) \tag{11}$$

$$M_2(\theta) = A(\theta) - P(\theta) \tag{12}$$

$$A(\theta) = \frac{M1(\theta) + M2(\theta)}{2} \tag{13}$$

$$P(\theta) = \frac{M1(\theta) - M2(\theta)}{2} \tag{14}$$

A second probe, orthogonal to the first one must be set if 2D knowledge of the positioning part (parasitic movement of a spindle) is required.

Note that this method is strictly similar to the layout with an invariant point defined by the probe as shown in Fig. 47, index 2: the difference is the fact that both artefact and measuring are now reversed and not the positioning part.

However, these conditions are sometimes impossible to get, only the artefact can be rotated, or even neither part. In addition the remounting operation of the artefact may induce additional errors. Finally, the parasitic movement of the positioning part may be not perfectly repeatable: the curve of the trajectory is not exactly identical over several turns, creating a band instead of a unique curve. The mean shape of the band corresponds to the fixed or synchronous component of the error and its width corresponds to the variable or asynchronous component of the error. In these cases, using multiprobe or multistep methods is necessary.

### 8.5 Fourier analysis and generalized diameter

Let us introduce the Fourier analysis since the methods exposed below induce harmonic losses in the results of assessments. Several probes set around the artefact constitute the measuring part of the system, whatever the type of the delivered information. The probe output is the sum of artefact and positioning errors for the $N$ measured points as in Donaldson's method. One can assume that the artefact shape or error curve can be described as a sum of harmonics, $n$ being integer and $N$ is the number of measurement points:

$$A(\theta) \approx \sum_{n=0}^{N-1} C_n e^{jn\theta} = \sum_{n=0}^{N-1} \left( A_n \cos(n\theta) + B_n \sin(n\theta) \right) = A_0 + \sum_{n=1}^{N-1} \left( A_n \cos(n\theta) + B_n \sin(n\theta) \right) \tag{15}$$

$A_n$, $B_n$, being the Fourier coefficients of the $n^{\text{th}}$ harmonic.

Consequently, the artefact error curve is a combination of basic shapes whose values $n = 3$ and $n = 4$ are shown in Fig. 51. The harmonic $n = 0$ is a constant and corresponds to the mean radius of the artefact: $A_0 = R_0$. All harmonics are centred on zero, the sum of any, over $\theta$ is null:

$$\sum_{\theta=0}^{2\pi}\left(A_n \cos(n\theta) + B_n \sin(n\theta)\right) = 0, \left(\forall n \neq 0\right) \quad (16)$$

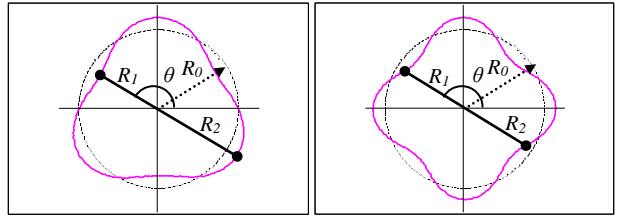

**Fig. 51:** Basic shapes for harmonics $n = 3$ and $n = 4$ and the corresponding generalized diameter for two probes: $D_{3/2}(\theta) = $ Const, $D_{4/2}(\theta)$, varies as the shape

A generalized definition of the diameter of a circular figure is useful for what follows. The diameter of a circle can be written $D = 2R$, but it is also $D = R_1 + R_2 = $ Const with $R_1 = R_2$, $\forall\ \theta$.

Applying that definition to the first harmonic of the artefact decomposition (Fig. 51) leads to $D(\theta) = R_1(\theta) + R_2(\theta) = $ Const in the case of a circle. A simple chart will show that it is still true for any odd-order shape. Conversely, the even-order shapes present a diameter varying with $\theta$ as the shape.

Suppose now that there are three probes equally spaced at $k2\pi/3$ around the artefact. The relationship $D(\theta) = R_1(\theta) + R_2(\theta) + R_3(\theta) = $ Const $\forall \theta$ is true for any $n$ shapes with $n \neq 3k$ (Fig. 52) and $D \neq $ Const if $n = 3k$.

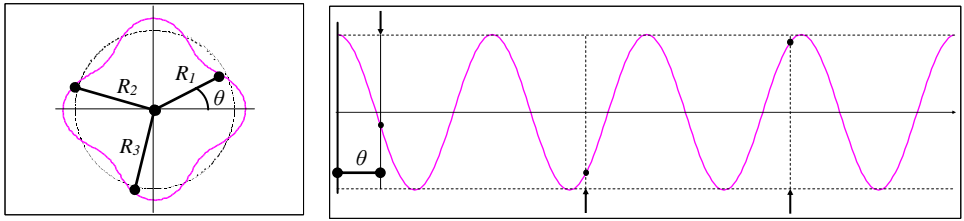

**Fig. 52:** $D_{4/3}(\theta) = $ Const, $\forall \theta$

Let the generalized diameter (GD) of an $n$-symmetrical figure and for $M$ equally spaced directions be:

$$D_{n/M}(\theta) = \sum_{m=1}^{M} R_m(\theta) . \quad (17)$$

The main interesting properties of the GD are

$$D_{n/M}(\theta) = \text{Const } \forall \theta \text{ for } n \neq kM \quad (18)$$

$$\frac{1}{M} D_{n/M}(\theta) \text{ varies as the shape itself for } n = kM \quad (19)$$

$$\frac{1}{2\pi}\sum_{\theta=0}^{2\pi} D_{n/M}(\theta) = R_0, \forall(n,M) \quad (20)$$

If necessary, and considering the artefact roundness being digitized by *N* points of measurements from the probes, the angle $\theta$ will be replaced by the number $i: 0 \to N-1$ where $\theta = i2\pi/N$.

$$\frac{1}{N}\sum_{i=0}^{N-1} D_{n/m}(i) = R_0, \forall(n,m) \quad (21)$$

Note that since the GD describes an error function, here of roundness, it can also be applied to an error function of tangential information. No particular condition concerning radial information has been expressed.

The concept of generalized diameter cannot replace Fourier analysis, but just offers a model for a qualitative approach. A thorough description of the analytic tools used in circular measurements has been synthesized by R. Probst in the appendix to his paper [9].

### 8.6 The multiprobe method (MP$^+$)

Constituting the average of evenly spaced probe outputs leads to the assessment of the variations of the GD of the artefact. But it cannot measure harmonics different from $n = kM, k \geq 0$.

Let us now try to manage the positioning part errors due to the spindle rotation. A displacement vector of the spindle rotor is seen by all the probes by the cosine for radial probes and by the sine for tangential probes of the angle $\varphi_m - \alpha(\theta)$ (Fig. 53). But the sum of all its contributions is always equal to zero at each step of $\theta$ because

$$\sum_{m=1}^{M}\cos(\varphi_m - \alpha(\theta)) = 0; \quad \sum_{m=1}^{M}\sin(\varphi_m - \alpha(\theta)) = 0, \quad (22)$$

with $\varphi_m = 2\pi m/M$

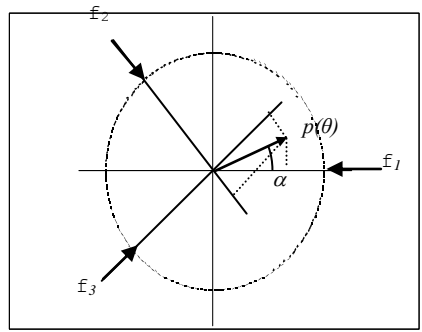

**Fig. 53:** Effect of a displacement vector $p(\theta)$ of the positioning part

In other words, the positioning errors are invisible on the GD of the artefact when averaging the probes, outputs, and the Fourier transform of the transfer function of the multiprobe method as in Fig. 54. Then, an assessment of *A* is

$$A(\theta) \approx \frac{1}{M} D_{n/M}(\theta) \approx \frac{1}{M} \sum_{m=1}^{M} M_m(\theta) ,\qquad(23)$$

with:

$$\frac{1}{N} \sum_{i=0}^{N-1} A(i) = A_0 = R_0 .\qquad(24)$$

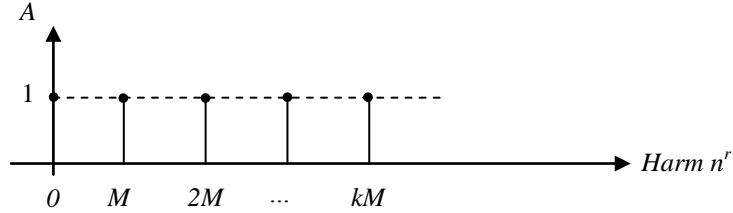

**Fig. 54:** Fourier transform (amplitude) of the MP$^+$ method with $n$ evenly spaced probes transfer function for the artefact roundness error

### 8.7 The multiprobe method with asymmetrical layout

The basic idea is to avoid the symmetry of the probes' angles since it is the origin of the harmonic losses. The probes are set at any $\varphi_m$ angles (Fig. 55). Using the GD extended to that kind of layout, it is clear that the condition $D_{n/M}(\theta) \neq$ Const is easier to reach than with the equally spaced probes.

A clever choice of the $\varphi_m$ can reduce drastically the harmonic losses in rejecting them far in the high frequencies (take care of which band is of interest).

However, the condition 'forced to be zero' on the positioning part is still necessary. Therefore a weighting GD is defined depending on the following conditions on the $\varphi_m$ angles:

$$\sum_{m=1}^{M} a_m \cos(\varphi_m - \alpha(\theta)) = 0 ;\quad \sum_{m=1}^{M} a_m \sin(\varphi_m - \alpha(\theta)) = 0 .$$

The GD (its variations) is used by summing the probes' outputs $M_m$:

$$D_{n/M}(\theta) = \sum_{m=1}^{M} a_m . M_m(\theta) .$$

Since the weighting coefficients distort the GD, it is necessary to correct its harmonic content. See more details about the method in D. Martin's lecture [10].

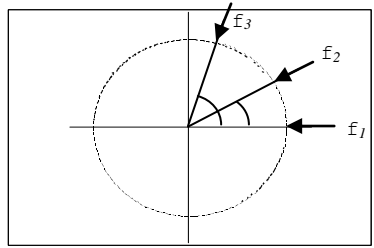

**Fig. 55:** The multiprobe method layout with asymmetrical angles

## 8.8 The multistep method (MS⁺)

The multistep method involves only one probe and, therefore, only one sensitive direction as in the Donaldson method. Since a rotation of $\pi$ rad of the artefact is not a true reversal, the idea is to rotate it $M$ times with an increment of $2\pi/M$ between sequences of $N$ measurement points (Fig. 56). Consequently, the artefact roundness error is shifted by $\varphi = 2\pi m/M$ at each sequence. The probe measures

$$M_m(\theta) = A(\theta - m\varphi) + P(\theta), \quad m = 1 \to M . \tag{25}$$

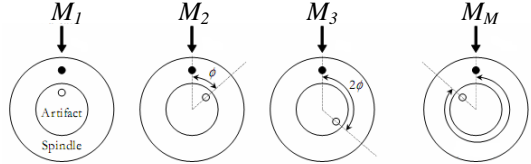

**Fig. 56:** Schematic of multistep method [11]

The example of the gear wheel measured by the CMM is actually a multistep layout with $M = N$.

The step $\varphi = 2\pi m/M$ being constant, the GD can be used: a radius at each step is measured. The GD (its variations) is calculated by adding the $M_m(\theta)$ together:

$$D_{n/M}(\theta) = \sum_{m=1}^{M} R_m(\theta) = \sum_{m=1}^{M} M_m(\theta) .$$

As for the multiprobe method, $D_{n/M}$ = Const for all the harmonics such as $n \neq kM$. On the other hand, the error due to the positioning part is measured $M$ times in the sensitive direction and in the same position (no rotation applied on the spindle). Then, constituting the GD as above and dividing it by $M$ gives the positioning part for all harmonics except for $n = kM$ for which it is not possible to separate from the artefact roundness error:

$$P(\theta) = \frac{1}{M} \sum_{m=1}^{M} M_m(\theta) . \tag{26}$$

The artefact error function is obtained by subtracting $P(\theta)$ from $M_1(\theta)$:

$$A(\theta) = M_1(\theta) - P(\theta) . \tag{27}$$

The Fourier transform of the transfer function for the multistep method is shown in Fig. 57.

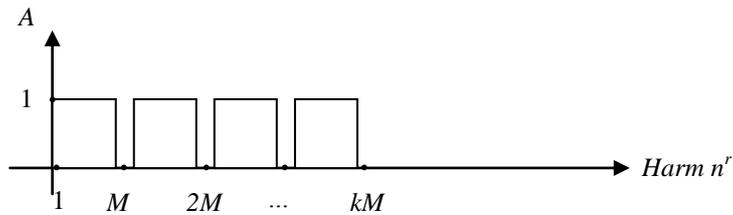

**Fig. 57:** Fourier transform (amplitude) of the MS⁺ method transfer function of the artefact with $\varphi = 2\pi m/M$ increment. The harmonic $n = 1$ is chosen equal to zero.

The eccentricity of the artefact related to the spindle induces the same kind of uncertainty on the harmonic $n = 1$ as for the multiprobe method.

Note that the multiprobe and multistep methods are fully complementary [12], adding both results leads to a true reversal (except for $n = 1$).

### 8.9 Errors due to instrument set-up in roundness or spindle error measurements

The relative positions between measuring and positioning parts and artefact mislead about the parasitic amplitude on some harmonics. The case seen above is the eccentricity between artefact and spindle whose influence is on the harmonic $n = 1$. Note that it is possible to calculate it (if it is only the origin of the first harmonic) by using directly the Fourier coefficients $dx = A_1$, $dy = -B_1$.

The tilt of the artefact (parallelism error between the plane containing the $N$ points of the artefact to be measured and the one containing the probes) can also influence the harmonic $n = 2$. But its influence is generally less sensitive than the effect of the eccentricity. The instrument set-up errors ($n = 1$, $n = 2$) cannot be separate from artefact or positioning parts.

### 8.10 Angular encoders: negative ESL and tangential measurements

Angular encoders are made of a glass circle whose circumference is accurately engraved up to 700 000 graduations (Fig. 42). The diameter of the circle may reach 140 mm.

The layout is the same as for the roundness error of an artefact. The only change is the type of information on the artefact which is tangential instead of radial. As for the radial measurement, the generalized diameter can be defined for the artefact error function. The positioning part of the system still exists with the two components $X(\theta)$, $Y(\theta)$, the read-heads correspond to the probes of the measuring part. They record the nominal value of an angle $\theta$, the error due to graduation defects, the one due to the positioning part between probes and circle and instrument set-up influence (Fig. 58):

$$M(\theta) = \theta + A(\theta) + P(\theta) + I(\theta) \ . \tag{28}$$

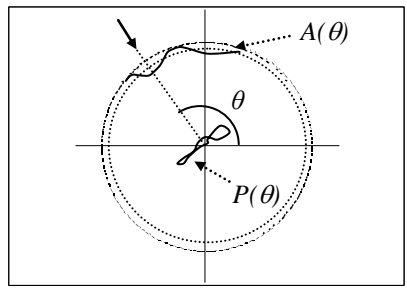

**Fig. 58:** Tangential measurements layout, $I(\theta)$ not shown

The following fundamental changes have to be noticed:

- The information goes from the artefact part (the circle) to the measuring one (the read-heads), contrary to the roundness error measurements (Fig. 59). This last remark is of the greatest importance because the transfer functions of the ESL methods do not have the same significance when using that layout for angular measurements: the transfer function of the multiprobe method with equally spaced probes appears to be very poor for roundness error measurement (Fig. 54). Since the roles are now inverted, that transfer function can be interesting for angular measurements to reduce the effect of the graduation errors on the angle assessment.

- The use of the system is then inverted: in the previous examples of roundness measurements, rotating the artefact over a turn was necessary for the assessment of the whole error function. When using angular encoders, the error function is not known and it is necessary to reduce its effect: an angle measurement is obtained by reading one time on the circle, this latter does not rotate over a turn. This is a negative ESL. The Fourier decomposition of the curve error still remains applicable even for unitary measurements (Fig. 60).
- The average of the error function of the graduations over $\theta$ is null: there is no concept of 'radius' anymore. Then, from Eq. (21):

$$\frac{1}{N}\sum_{i=0}^{N-1} A(i) = A_0 = 0 \ . \tag{29}$$

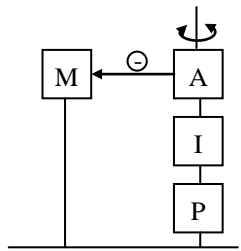

**Fig. 59:** The ESL model of angular encoder

A layout with two opposite probes (Fig. 60) using the average of the read-heads outputs (MP2⁻) eliminates errors due to eccentricity between the circle (A) and probe support (M), all the odd harmonic errors of graduations, and finally the positioning part variations since it cannot be seen by the MP layout [(Eq. 22)].

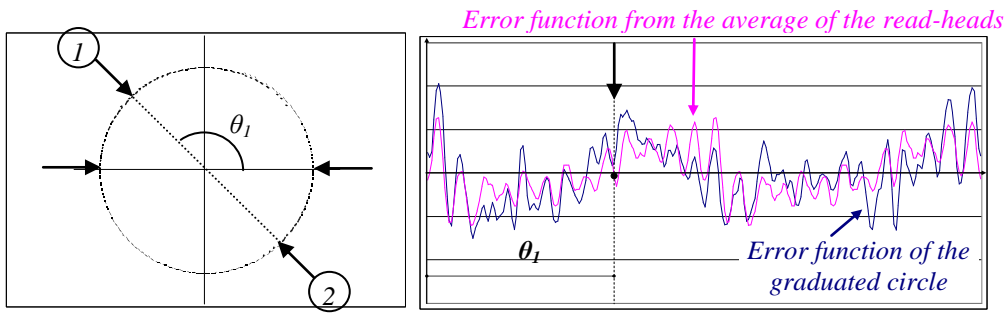

**Fig. 60:** $\theta_1$ angle obtained from a pair of opposite read-heads (MP2⁻)

The four-probes configuration (MP4⁻, $\varphi = k\pi/2$) is shown in Fig. 61. Both eccentricity ($n = 1$) and tilt ($n = 2$) due to instrument set-up are eliminated with the average of the outputs. Therefore, only $A(\theta)$ at $n = 4k$ harmonics affect the assessment of the angle $\theta$.

We insist on the fact that the benefit from the MP transfer function on angular measurements is only due to the inversion of the information direction and not to its tangential nature.

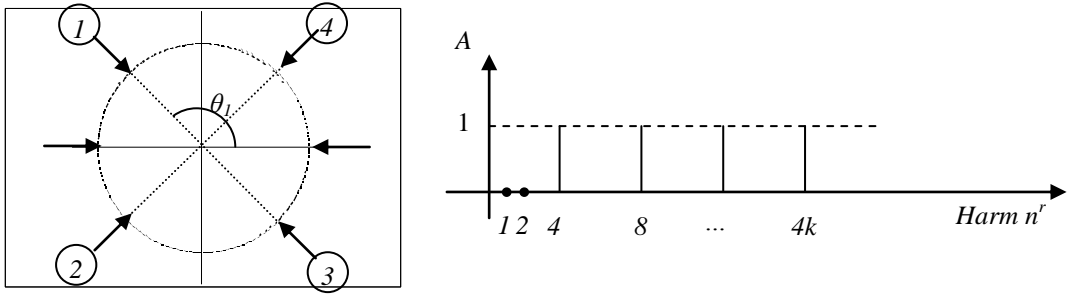

**Fig. 61:** Fourier transform (amplitude) of the transfer function for four read-heads of an angular encoder (MP4$^-$): $n = 1$ (eccentricity) and $n = 2$ (tilt) are eliminated

### 8.11 Continuous measurement with angular encoders

A first approach for reducing the error function would consist in an increase in the number of probes. But we quickly face a limit due to cost and the physical layout. The second one consists in rotating the circle over a turn. The probes register the whole error function whose average is null [(Eq. 29)]:

$$\sum_{i=0}^{N-1} M(i) = N\theta + \frac{1}{N}\sum_{i=0}^{N-1} A(i) = N\theta \quad \Rightarrow \quad \theta = \frac{1}{N}\sum_{i=0}^{N-1} M(i) \ . \tag{30}$$

In order to cancel the instrument set-up errors, the principle must be applied with the MP2$^-$ or MP4$^-$ layouts of angular encoders. The term $M(i)$ in Eq. (30) is replaced by the average of a MP layout.

An angle $\alpha$ as shown in Fig. 62 is computed with the difference of two MP2$^-$ layouts: the circle rotates over $\theta$ in the MP2$^-_1$ configuration, then the measuring part supporting the two read-heads is rotated by $\alpha$ and the circle rotates again over $\theta$ in the MP2$^-_2$ configuration. The angle assessment is

$$\alpha = \frac{1}{N}\sum_{i=0}^{N-1}\left[MP2^-_2(i) - MP2^-_1(i)\right] \ . \tag{31}$$

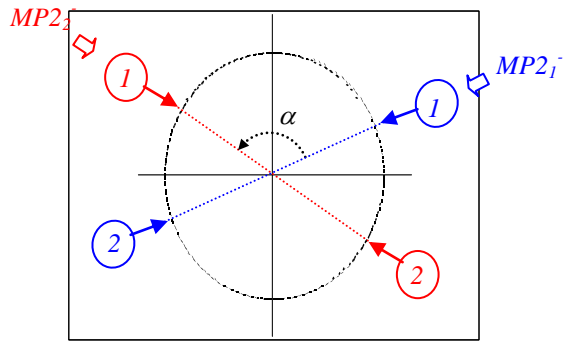

**Fig. 62:** An angle measurement without any error of graduated circle

That is why the dynamic sensors of accurate theodolites were introduced around 1980. The Wild T2000 has a circle graduated with only 1024 divisions [13]. The circle describes a full revolution for each angle measurement and is scanned by two read-heads, one fixed as a reference and the other mobile and mechanically linked to the telescope (Fig. 63). The phase shift between the two signals makes it possible to calculate the angle formed by the two read-heads. It is the average value of all phase shifts which gives the precise measurement.

Actually, the T2000 is equipped with two pairs of opposite heads to eliminate the variable effect of the eccentricity (not shown in Fig. 63).

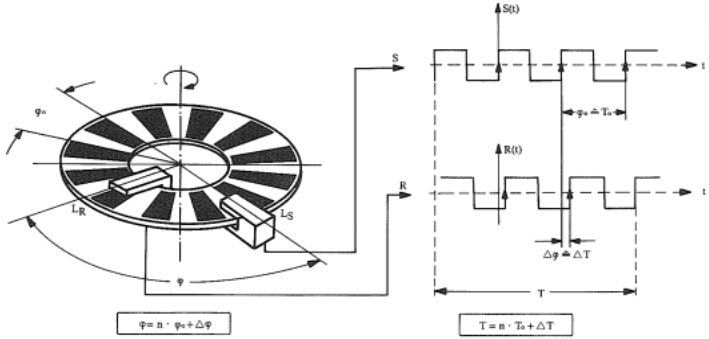

**Fig. 63:** Dynamic angular encoder of the Wild T2000

Similar assemblies exist for the realization of rotating plates of very high degree of accuracy (few 0.01") [4], [10], and [14]. The principle of the ESL is the same as for the T2000. Such rotating plates differ because they are equipped with a set of two encoders (circle + read-heads) mechanically mounted in juxtaposition to each other (Fig. 64). The circles are interdependent, turn together, and constitute part $A$ of the ESL, each encoder (set of read-heads) being the $MP^-_1$ and $MP^-_2$ layouts of the probes.

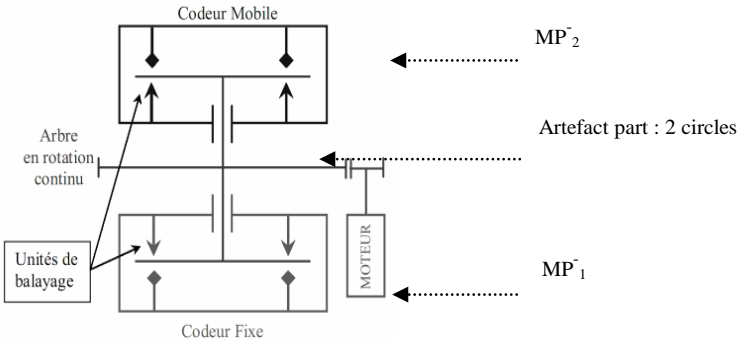

**Fig. 64:** Double encoder [4]

## 8.12 Calibration of graduated circles

Another use of the rotating circle of angular encoders is very similar to roundness error assessment of an artefact: it is sometimes necessary to assess the error function of the graduations in order to use it for checking manufacturing or for an *a posteriori* calibration use. In that way, the ESL problem (whatever the approach for calculation) is exactly the same as for the roundness measurements; the sense of measurement is direct ($MP^+$, $MS^+$): the wider range of the transfer function has to be found. In the so-called equal-division-averaged (EDA) method, two circles are compared and both layouts are used in parallel: $MP^-$ for the reference circle and $MP^+$ for the circle to be calibrated [15].

## 8.13 Multistep layout for angular measurements

Measuring an angle $\alpha$ with the multistep method is possible. The errors of the circle are reduced if the angle measurement is iterated with an increment of 360°/$M$ of the circle: for $M = 4$ (Fig. 65), the engraved circle is rotated by $\varphi = \pi/4$ before a set of measurements is carried out in the two directions defining the angle α.

One can prove that using the average of a pair of opposite read-heads and taking the $M$ averages corresponding to the $M$ steps, for calculating the GD as for the radial measurements [(Eq. 26)], gives exactly the same transfer function as for the multiprobe layout (Fig. 57). In this case, Eq. (27) is not used anymore.

But the multistep method assumes that the positioning part and the instrument set-up errors are synchronous. This assumption is not enough for very precise encoders. That is why the multiprobe layout is always preferred for high-precision angular encoders.

Before the advent of the current modern theodolites, geodetic networks campaigns could include up to 16 iterations for an instrument equipped with a pair of opposite read-heads [16].

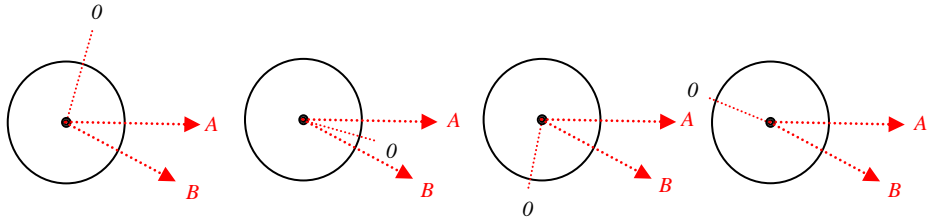

**Fig. 65:** Theodolite measurement in MS4⁻ layout

## 8.14 Matrix-based approach

Another way to solve the ESL problems is to define a set of linear equations describing the layout. The first step is to detect the unknowns involved in the layout. As an example, a MP3⁺ configuration with $N$ measured points is described by $N$ unknowns $A(\theta)$ for the error function of the artefact and $2N$ unknowns for the 2D curve of the positioning part: $X(\theta)$, $Y(\theta)$. On the other hand, the set of three probes gives $3N$ data. The set of equations does not allow redundancy in this example but a unique solution exists.

In the MP3⁺ method of radial measurements, the following equations describe the system shown in Fig. 66:

$$A(\theta) + P(\theta) = M(\theta) \Leftrightarrow N.X = M \qquad (32)$$

where 
$$A = \begin{bmatrix} Id & Id & 0 \\ Id & \cos(\varphi_2).(Id - \varphi_2) & \sin(\varphi_2).(Id - \varphi_2) \\ Id & \cos(\varphi_3).(Id - \varphi_3) & \sin(\varphi_3).(Id - \varphi_3) \end{bmatrix} \text{ with dim}(A)=3N. \qquad (33)$$

$M$ is a vector built with the probes outputs. $X$ is the vector of the unknowns. $Id$ is the matrix $Identity$. $Id - \varphi_i$ are matrices built with the circular permutation of the columns of $Id$ by a number corresponding to $int\left(\dfrac{\varphi i}{2\pi}.N\right)$.

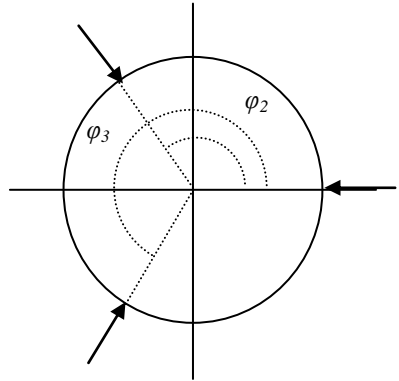

**Fig. 66:** MP3$^+$ at k2π/3 layout

The general solution is obtained by solving the following equation according to the least-squares principle:

$$X = (A^T.A)^{-1}.A^T.M \quad . \tag{34}$$

In the case of a MP3 with three equally spaced probes, the result of matrix-based calculation is the **exact** solution (except for the harmonics involved in the instrument set-up) providing the data are also exact, i.e., without any noise. Conversely, the computation based on the GD always presents losses (see Section 8.5). However, the result may suffer from the same problem, depending on the level of noise on the measurement. Simulations show that the amplitude of losses is often lower in matrix-based computation than in the GD approach in the presence of white noise on the probe outputs but they also show that the accurate knowledge of the $\varphi_i$ angles of the probe location is then more sensitive.

Since the system is rank-deficient, because that kind of matrix shows Det*(N) = 0*, inverting the matrix *N* leads to instabilities. Another way, which is far better, is to use the pseudo-inverse matrix of *N*. The equation is then

$$X = \text{pinv}(N).M \tag{35}$$

Figure 67 shows an example of the comparison on a set of data in a MP3$^+$ layout of simulated roundness-error measurements and calculated with ™Matlab and using the ***pinv*** function. Here are the corresponding layout parameters:

– MP3$^+$ layout

– $\varphi_i = k2\pi/3$

– $A(\theta)$ is a broadband signal.

– $N = 256$ measurements

– Any $P(\theta)$ is used

– The level of noise applied to the probe outputs is less than 1% of the $A(\theta)$ amplitude

– No $d\varphi_i$ is applied here

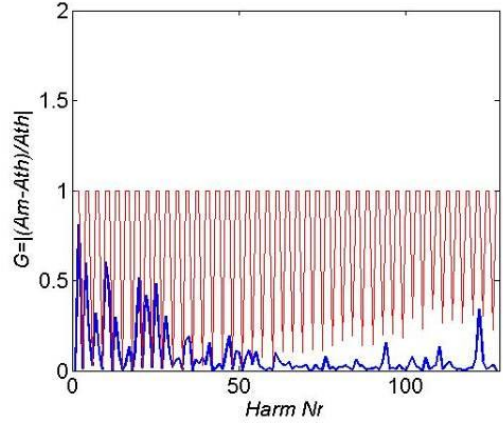

**Fig. 67:** Comparison of GD- and matrix-based calculations of the artefact for a symmetric MP3$^+$ layout: noise on output probe effect

The error function gain is calculated for both solutions: $G = (A_{mes} - A_{theo})/A_{theo}$ where the $A_i$ are the amplitudes of the harmonics. $G = 0$ means that the harmonic is fully determined, $G = 1$ means its amplitude is null. The theoretical Fourier transform of symmetric MP3$^+$ appears clearly on the GD-based curve (red line) and the comparison is in favour of the matrix-based calculation (blue line). These results must be interpreted as the ultimate capability of the matrix-based approach because the effect of the uncertainty on the true location of the probes of about $d\varphi_i = 2\pi/N$ may be unacceptable (Fig. 68).

The proposed example does not lead to very good results when using it to assess the roundness; the asymmetric MP3$^+$ is better. In that last case, GD- and matrix-based approaches tend to have the same behaviour in presence of a $d\varphi_i$ shift.

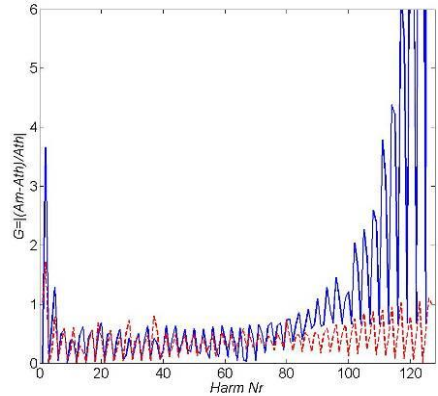

**Fig. 68:** Comparison of GD- and matrix-based calculations of the artefact for a symmetric MP3$^+$ layout: noise on output probe + $d\varphi_i$ shift effect

The matrix-based approach in ESL can be delicate. The pseudo-inverse and the Singular Value Decomposition (SVD) may be interesting ways. Note that some authors have already proposed a novel approach with Prony decomposition involving SVD without harmonic losses [17].

In any case, applying matrix calculation for angular encoders is not necessary; the GD-based approach with engraved circle fulfils the challenge of precision, with or without rotation of the circle.

## 8.15 The ESL approach in magnetic measurements

Rotating coils are intensively used in the field of particle accelerators for the magnetic characterization of the magnets. The level of accuracy in terms of dimensional metrology can reach 0.01mm and 10 ppm in terms of magnetic parameters. That is why some care must be taken in the design and the procedure of magnetic measurements.

Since the bench dedicated to this use often includes a rotation part, coils of Hall probes, it is possible to find a correspondence between ESL and the techniques of the magnetician. The basic equation describing the magnetic field in a multipole magnet is as follows:

$$B(z) = \sum_{n=1}^{N(=\infty)} C_n \cdot \left(\frac{z}{R_r}\right)^{n-1}, \quad (36)$$

where $z = x + jy = re^{j\theta}$ is the affix of any point located in the field and with $C_n = B_n + jA_n$.

Here $B_n$, $A_n$ are the harmonic coefficients obtained by Fourier transform[1]. $R_r$ is an arbitrary reference radius acting as a normalization of the calculations.

One has to remember that the magnets we deal with are $2m$-poles. The corresponding field harmonics are $m = 2n$ (Fig. 69).

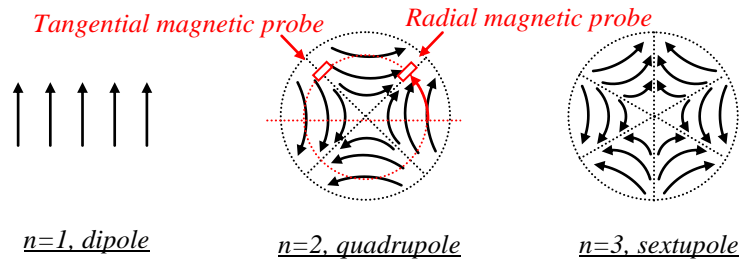

*Tangential magnetic probe*   *Radial magnetic probe*

n=1, dipole     n=2, quadrupole     n=3, sextupole

**Fig. 69:** The three first harmonics of $2m$-pole magnets

The magnetic probes can run tangentially or radially. The set 'probe-support' is known as the detector. Both rotate as shown in Fig. 69 (red lines on Qpole chart), measuring the flux $\phi(r, \theta) = \phi(\theta)$, $0 < \theta < 2\pi$, since $r$ = Const allows the calculations of the harmonic coefficients of the field $B(z)$ through its Fourier transform.

Even if the typical Fourier transform of the multipole magnet fields is simple (only one main harmonic, the fundamental corresponding to the kind of the magnet), the undesired harmonics to be measured are less than $10^{-4}$ of the fundamental.

In terms of ESL, one can define the artefact part as the magnetic field or, rather, the flux measured by the probe. It corresponds to a signal being a function of the probe rotation $\theta$. The measuring part corresponds to the probe (one or more) and the positioning part to the synchronous parasitic displacement of the probe when rotating. In that layout, the $P$ part is applied to the measuring part instead of the artefact. Figure 70 shows the proposed model compared to the one of roundness measurements.

---

[*]In magnetism, the coefficients $A_i$ and $B_i$ are inverted with respect to the introduction to Fourier analysis (Section 8.5).

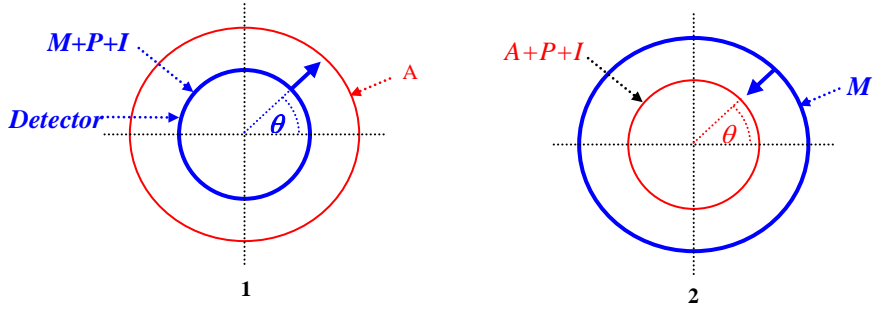

**Fig. 70:** ESL comparison: **1**: rotating magnetic probe; **2**: roundness measurements

That arrangement with a detector of only one probe can measure all the harmonics of the field (MP1$^+$ on Fig. 71). It is a simple version which is actually very common. There are $N$ unknowns of the artefact part defined by an increment of $\theta = 2\pi/N$. Another ESL is present on the bench with the use of a two-read-heads angular encoder and linked to the detector shaft (MP2 on Fig. 71) to control the rotation $\theta$.

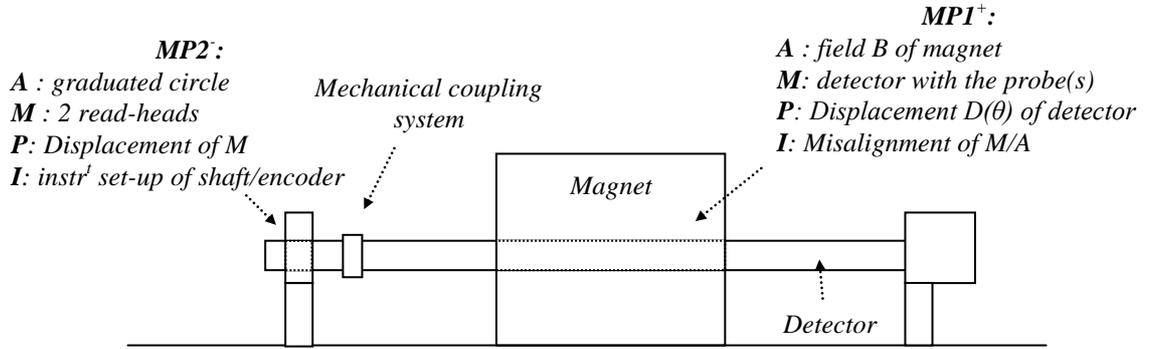

**Fig. 71:** The ESL on a bench for magnetic measurements

Instrument set-up errors also exist in such magnetic benches: the flux measurements are affected by the misalignment of the detector with respect to the centre of the magnet field. If the offset between the centre $z_c$ of the magnet, typically a quadrupole, and the centre $z_m$ of the detector rotation is $d = (z_c - z_m)$, then every Fourier coefficient $C_{n_{meas}}$ measured by the detector is a function of

$$\sum_{k=n}^{\infty} C_k \frac{(k-1)!}{(k-n)!(n-1)!} \left( \frac{d}{R_{ref}} \right)^{k-n} \quad \text{(Ref. [18])} .$$

That function highlights the nature of the artefact which is different from dimensional radial or tangential measurements: an eccentricity modifies only the first harmonic in the dimensional domain but all of them in magnetism. It is due to the artefact in magnetism (field $\vec{B}$) whose harmonic spectrum varies as $r^{n-1}$ instead of $r^1$ in the dimensional domain. However, applying the formula to $n = 2$ corresponds to the dimensional metrology case, and also to the quadrupole field. The formula gives

$$dx \approx \frac{R_r}{2} \frac{B_1}{B_2}, \qquad dy \approx -\frac{R_r}{2} \frac{A_1}{B_2} . \tag{37}$$

Therefore, the first harmonic is supposed to be due to the misalignment offset only, and the measured coefficients $B_1$ and $A_1$ are then used to give the co-ordinates of ($z_c$ - $z_m$) as in dimensional metrology (see Section 8.9).

Using the first harmonic to assess the misalignment as for the roundness error measurements assumes that there is no artefact information in it, i.e., no dipole field created by the quadrupole. That hypothesis is usually reliable.

The angular encoder may show a similar offset: it can be not coaxial with the shaft axis of the detector. The effect is of the same kind as between detector and quadrupole: the encoder delivers an output affected by an error $e_\theta \propto \sin\theta$, creating a harmonic $n = 1$ (dipole) in the measured spectrum of the field. The parallelism defect between encoder and shaft axes induces an error on the harmonic $n = 2$ similar to the one in ESL due to the tilt effect (see Section 8.9).

Even if the magnetic benches have excellent quality of manufacturing, especially for the shaft of the detector, parasitic displacements may remain during its rotation. The problem is well known in dimensional metrology; that is the positioning part of the ESL. It has been proved that the main part of the effect appears on the harmonic ($n = 1$) of the measured spectrum of a $2m$-pole magnet [18]: such displacements can be interpreted as an instantaneous misalignment of the detector, depending on $\theta$. The eccentricity $E$ is fixed over $\theta$, the displacement $D(\theta)$ is a function of $\theta$.

A possible improvement could be the design of a three radial coils detector with a dissymmetrical layout (MP3$^+$ dissymmetric, see Section 8.7) for the quadrupoles' magnetic measurements since they are quasi similar to the dimensional case and since their misalignment is sensitive. Two more coils could control the parasitic displacement of the M part during its rotation if required. The level of rejection to higher harmonics that shows the dissymmetrical layout would be easily favourable as regards the harmonic distribution of quadrupolar field. That proposal has to be tested to be confirmed, especially because the probes in magnetism measure the flux and not the field and therefore its harmonic spectrum varies as $r^n$. The probes do not run like in dimensional measurements, even for a quadrupolar magnet:

$$\Phi(\theta) \propto \int_r B(r,\theta)dr = f(r^n) \ [18] \ .$$

The approach used so far for the magnetic measurements ESL analysis can be completed by the following examples of what could be a parallel to ESL in that domain. It is sometimes necessary to measure only an harmonic, say $n = 5$, corresponding to a decapole component of the field. A solution is to use a detector with five probes (coils or Hall probes) equispaced with $\varphi_i = k2\pi/5$ (Fig. 72) and the harmonics $n = 5k$ are measured. Such a detector has been designed in the LHC project for static measurements (no rotation of the detector) [19] but it is sometimes used as a MP5$^+$ arrangement with rotation. The same is true with the use of two identical coils in electric opposition [20]: the odd harmonics are rejected, especially the one of the dipole ($n = 1$). In these two examples, the positioning part of ESL is supposed to be null; it is not a true error separation situation.

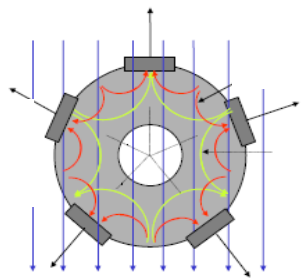

**Fig. 72:** Five equispaced Hall probes ⇔ MP5$^+$ [17]

## 8.16 Conclusions

The GD-based approach cannot replace the Fourier analysis tools which are far more complete. But what we want to do is to enlarge the model qualitatively. The ESL model could be a synthetic point of view common to many applications in very different domains such as surveying, mechanical metrology or magnetism. Keeping in mind the ESL model would help any designer or metrologist in any domain of the measure. Further investigations would be of great interest for applying ESL in the domain of rotating coils.

# 9 Differential measurements

The principle of differential measurement is extremely common in all fields of metrology. Instead of measuring a distance between two points with a ruler, by applying an extremity to a point and by reading the graduation corresponding to the other one, the ruler is shifted and two readings on the ruler are carried out in front of the two points (Fig. 73). The length is the difference between the readings.

The first method, known as 'direct measurement', shows a regular problem due to the accessibility of the ruler's origin. The use of a measuring instrument often needs a material definition, a mechanical reference which is supposed to be representative of the origin of measurement itself. Unfortunately, a good precision of the zero adjustment with respect to a mechanical reference is seldom possible. Calculating the difference between two measurements eliminates it:

$$d = (L_2 + e) - (L_1 + e) = L_2 - L_1 \quad . \tag{38}$$

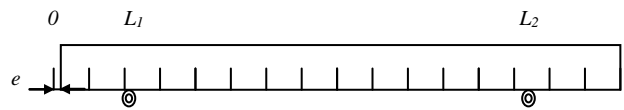

**Fig. 73:** Differential measurement

This example, rather commonplace in the field of length measurements, is evident in the angle measurements. Indeed, an angle measured with an angular encoder is inevitably a difference of two directions. However, that kind of differential measurement cannot correct the linearity errors of a measurement system, only a calibration can solve it.

The application of the principle can be considered in a 2D or 3D configuration, the example of the ruler or the angular sensor being in 1D.

## 9.1 Ecartometry measurements

A wire ecartometer is used to measure the misalignments of components by means of their fiducials. The instrument (Fig. 74), centred on the fiducials of the component to be aligned, measures the distance from this point to the line defined by a wire stretched between the fiducials at both ends (Fig. 76). The adjustment by the zero measurement in coincidence with the mechanical reference of the instrument is difficult to realize. Each reading $L_i$ is affected of the zero offset $e$ of the instrument: $L_i = D_i + e$. In addition, centring the wire precisely on the two fiducials of extremity is very expensive in terms of accuracy and offsets also exist on the wire position (Fig. 75). The elimination of these errors is carried out by measuring with respect to a wire stretched between two other points which are not the fiducials and by changing the base of calculation: thus one creates a fictitious wire and measures without offset.

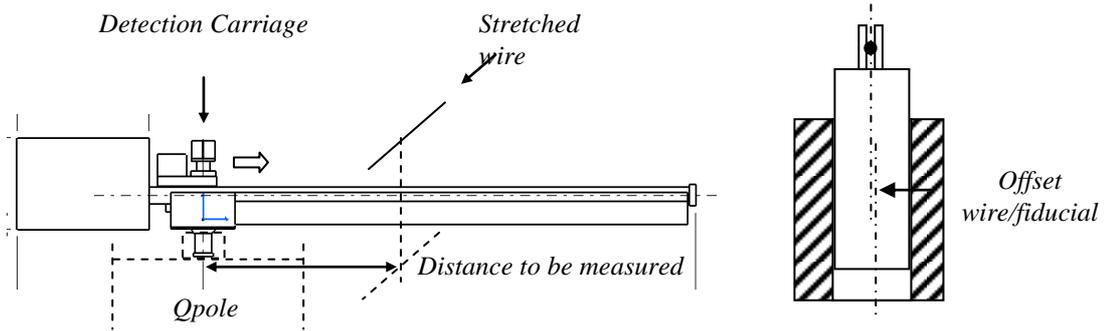

**Figs. 74, 75:** Wire ecartometer and centring error

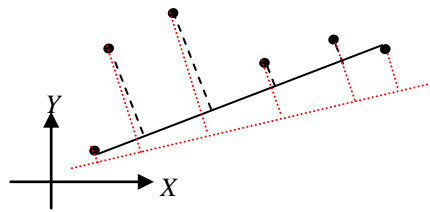

**Fig. 76:** Differential measurements in ecartometry

## 9.2  $V_0$ in geometry

The use of a tacheometer (measurements of both angles and distances) to determine the Cartesian co-ordinates of the points *C* and *D* while knowing those of the points *A* and *B* is simplified by adding a variable in calculations $V_0$ which corresponds to the direction of the zero of the angular encoder of the instrument (Fig. 77). The variable $V_0$ is calculated from the known co-ordinates of points *A* and *B*:

$$V_0 = V_{AB} - L_B .$$

Then, the directions of the vectors *AC* and *AD* are calculated with respect to the *y*-axis (direction) from the $V_0$: $V_i = V_0 - L_i$.

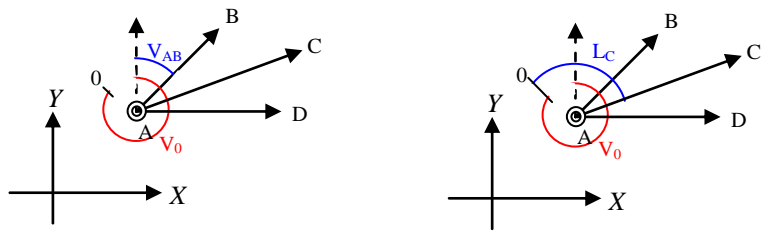

**Fig. 77:** $V_0$ in planimetry

The $V_0$ has the same role as the ecartometer offset *e* of the previous example. The zero of the circle is not known and justifies its use.

## 9.3 Network of a Hydrostatic Levelling System (HLS)

The HLS is a capacitive non-contact displacement sensor which runs with electric conductors and especially with free surface of water (FSW) defining a perfect horizontal plane (Fig. 78). It enables measurements and adjustments of relative position between structures in the vertical direction. When a hundredth of mm accuracy is required, the determination of the offset *e* of the sensor due to the zero drift with respect to its mechanical reference is critical. We present an example using as many sensors as the *N* points to be controlled (Fig. 79).

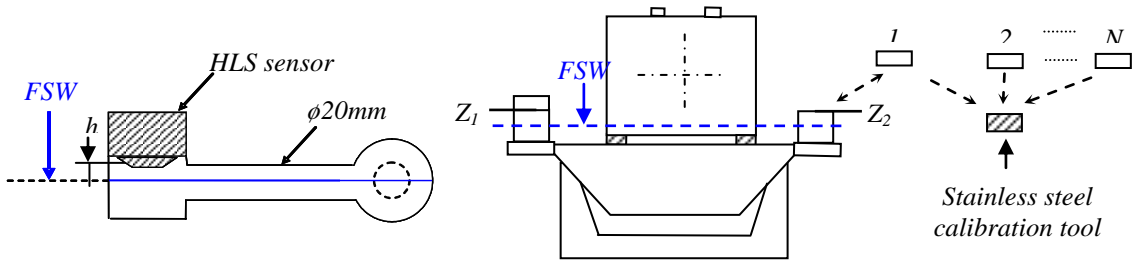

**Figs. 78, 79:** HLS sensor and HLS network and calibration tool

*N* sensor offsets are to be determined. The principle of differential measurements intervenes first, on the relative positions $dZ = Z_1 - Z_2$ of the zeros which are to be determined, not their true position *h* with respect to the surface of water, which only represents a precise reference of horizontality. The differences in the readings *dL* from the sensors give the position information *dZ*. But each sensor $HLS_i$ is affected with an error $e_i$. Since the configuration is differential, then the principle also applies to offsets. The problem is solved by the use of a stainless-steel calibration tool (Fig. 79) which is measured by all the sensors to compare their offsets: the common part of the offsets disappears: $h_1 - L_1 + e_1$; $h_2 - L_2 + e_2$; $de = e_1 - e_2$ is measured with the tool. Then $dZ = h_1 - h_2 = L_1 - L_2 + de$.

## 9.4 Interferometry

The interferometry principle is a very powerful tool in DM whatever the scale of the measurements. It is employed in microscopy and astronomy. What is interesting here is that the differential principle exists at the physical level. The fringes of two signals, one from the measuring channel and the other from the reference, are combined together. The scale of the phenomenon is given by the wavelength of the emitted light; it means that the sensitivity can be very important; the level of a nanometre is reached in many cases.

The most common arrangement in DM is achieved with the Michelson interferometer (Fig. 80) where a half-transparent mirror (beam-splitter) splits a laser source into two beams, one acting as a 'reference' and the other one being the 'measuring' one. The two beams are recombined, creating a pattern of fringes and then processed after each of them has been reflected onto a mirror.

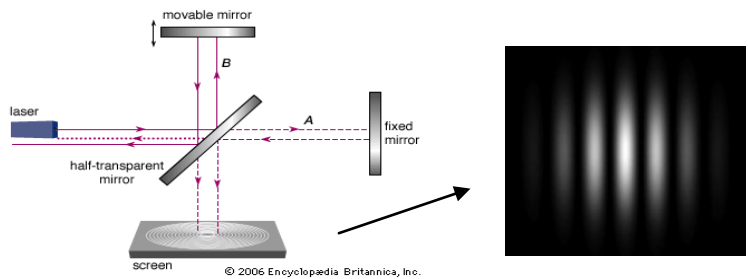

**Fig. 80:** Michelson interferometer applied to linear displacement measurement

If a mirror moves, typically the 'measuring' one, while the other mirror stays fixed, a fringe pattern scrolling arises, depending on the distance *d* of displacement. Counting the fringe leads to the distance *d*.

$S_1 = a_1 \cos(2\pi f t)$ is the signal coming from the reference, while $S_2 = a_2 \cos(2\pi f t + \varphi)$ is the one coming from the moving mirror. $\varphi$ is the phase difference due to the displacement:

$$\varphi = k \frac{2\pi}{\lambda} 2d \ . \tag{39}$$

$\lambda$ is the wavelength of the light source in the air. If $d = \lambda/2$ and $k = 1$, then $\varphi$ describes $2\pi$ in phase. That is why $\lambda/2$ is considered as the basic increment of such methods. Counting *k*, the number of fringes leads to the distance $d = k \frac{\lambda}{2}$.

The mirrors can be replaced by retro-reflectors which have the advantage of reflecting accurately the beam parallel to the incident one. However, flat mirrors are always used for ultimate accuracy.

Several models of interferometer exist. The heterodyne interferometer using two frequencies is very common in metrology laboratories (Fig. 81). The laser source includes two frequencies $f_a$ and $f_b$ with in quadrature polarities. The signal $f_b$ - $f_a$ - $\Delta f$ is analysed, where $\Delta f$ is the shift frequency due to the translation of the 'measuring' retro-reflector.

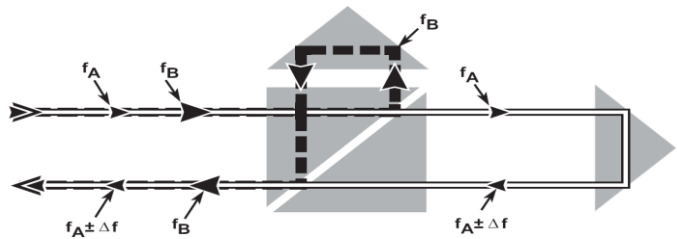

**Fig. 81:** Heterodyne interferometer: layout for linear measurements (™Agilent)

Interferometry can be applied in other domains of DM such as angle measurements or straightness measurements. Figure 82 shows the schematic of differential linear measurement applied to angles: the so-called 'reference' reflector of the linear set-up is now linked to the measurement one. Any rotation of the angular reflector induces a path difference between $f_a$ and $f_b$ corresponding to a relative displacement *d*. If the spacing *H* of the two retro-reflectors is accurately known, the angle displacement can be computed: $\sin(\alpha) = d/H$.

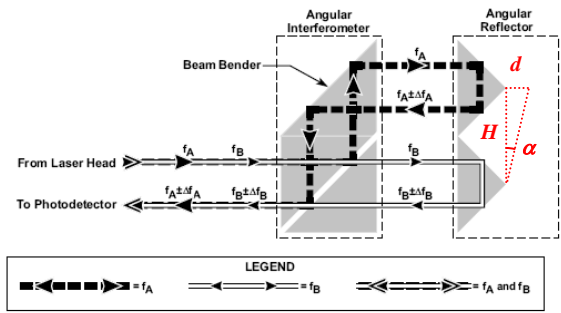

**Fig. 82:** Heterodyne interferometer for angular measurements (™Agilent)

In order to reach ultimate accuracy (few nm), the differential principle is applied at many levels. Thus more precise optics need to mix both frequencies on the whole optical path so that they 'see' exactly the same errors (Fig. 83).

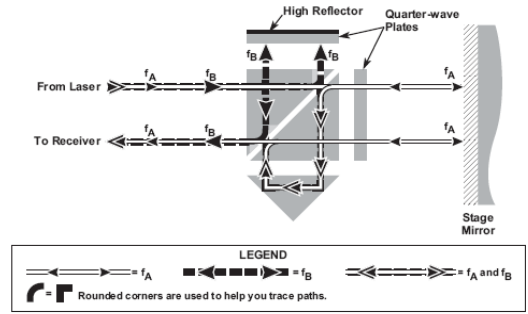

**Fig. 83:** High-accuracy optics for linear measurements (™Agilent)

## 9.5   Differential measurements with magnification: a link with reversal layout

Some layouts allow the measurement of the double of a physical quantity to be measured. These are typically the reversal layouts where the error is as $2e = R_1 - R_2$ (Section 7.1). Some physical arrangements can reach such effects. They also correspond to a reversal arrangement, the physical $\pi$ radians rotation being always present. The use of a retro-reflector shifts a beam offset (Fig. 84). Using a plane mirror for autocollimation is exactly the same situation: the sensitivity of the method is multiplied by two (Fig. 36). That principle is also applied for reading at opposite read-heads of an angular encoder [21] or viewing the extremities of a levelling machine bubble [Fig. 85(b)].

In the field of electronics, quad cells for photon beams as well as BPMs for electron beams run identically [(Fig. 85(c)]. The beam position variations in $z$, for instance, are computed from the four electrode signals of the BPM [22]:

$$z \propto \frac{(I_A + I_B) - (I_C + I_D)}{(I_A + I_B + I_C + I_D)} \ . \tag{40}$$

The sum of the current in the denominator is constant for small displacements of the beam and acts as normalization with respect to the beam current. But the structure of a reversal is still present even in Eq. (40). The four electrodes are used in the horizontal direction as well ($x$), and two symmetry axes exist for the $\pi$ radians rotation of the reversal layout.

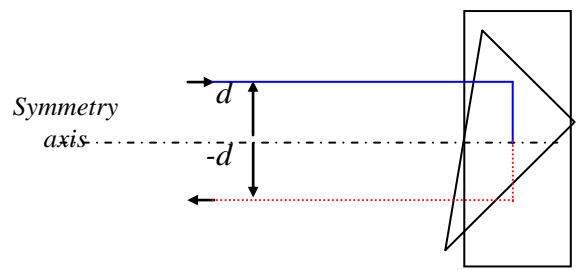

**Fig. 84:** Symmetry in a retro-reflector

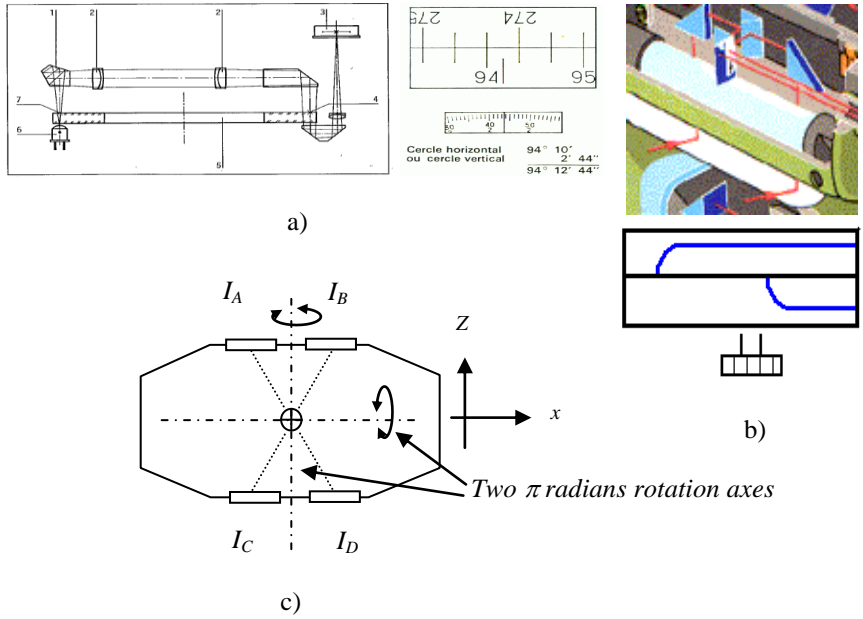

c)

**Fig. 85:** Instrument components showing differential function: (a) opposite read-heads of an optical angular encoder (theodolite), (b) bubble of a levelling machine, (c) BPM

Furthermore, magnification can be greater than two. The classical example of the use of the reticle pattern with two sloped lines aiming at a target is interesting. The pattern has an angle $\alpha$ and the upper and lower lines intersect the target differently. Figure 86 (which is drawn to scale) shows a vertical displacement $h$ of only 2% of the width graduation and induces a $dx$ shift of 28% corresponding to a magnification $G \cong 13$. The corresponding formula is

$$dx = 2h \cdot \cot(\alpha/2) \ . \tag{41}$$

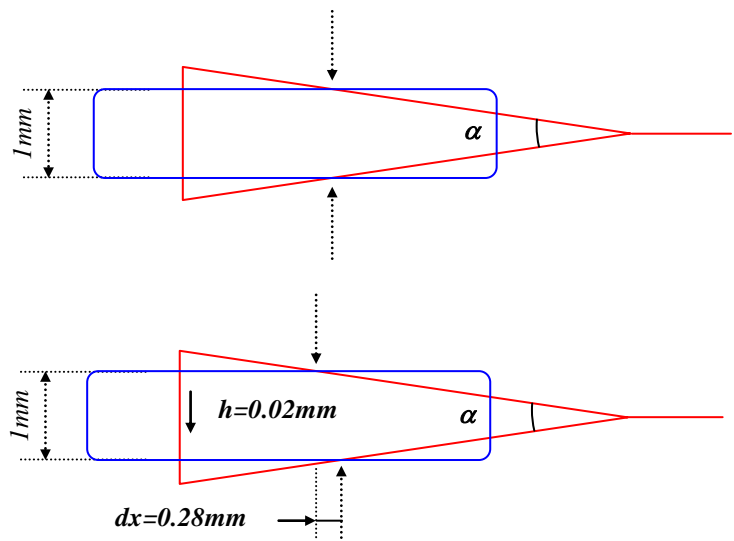

**Fig. 86:** Differential measurement with magnification (the proportions are correct))

# 10 Case study: the orbit definition of a synchrotron storage ring

## 10.1 Presentation

The beam orbit of a storage ring is fully defined by the location of its quadrupole magnets (Qpoles) (Fig. 87). The beam quality depends on the quality of the orbit. Before the alignment operation of these magnets, it is necessary to materialize their magnetic axis. In terms of metrology loop, it consists in linking the magnetic definition to the fiducialization. Then, magnets can be adjusted with each other by means of mechanical and/or optical methods. The magnetic axes of all the magnets are then linked together to define the orbit. The first step of the alignment is the magnetic axis detection of the Qpoles. A magnetic bench is dedicated to that operation (Figs. 87 and 93) from which two kinds of fiducials are adjusted (for mechanical alignment on girders) or measured (for optical adjustment in the tunnel). These latter are surveyed with a special structure called 'Qpole comparator' with respect to the cylindrical support of the rotating coil.

The Qpoles are then mounted on their girder and laser measurements are carried out to check their mechanical alignment.

The last step is the precise alignment of all Qpoles together to define the storage ring orbit with wire ecartometers for the planimetry and HLS for the altimetry.

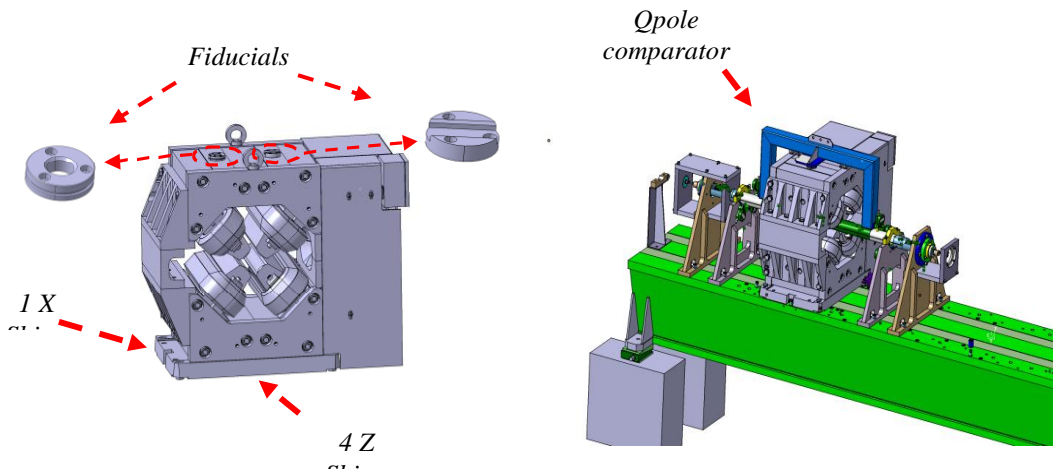

**Fig. 87:** Storage ring quadrupole at SOLEIL and bench for magnetic measurements

## 10.2 Detection with rotating coil

The description of the magnetic bench dedicated to the SOLEIL quadrupoles is given in detail in Arnaud Madur's thesis [22].

Rotating coil measurements allow knowledge of the magnet magnetic field **B**. The coil rotates and then measures the flux coming from the magnet. The coil is set radially or tangentially. In the present case study, a radial coil is used (Fig. 89). A Fourier transform is applied to the flux signal. The magnetic axis of the Qpole is then located (Fig. 88) by applying the following formulas:

$$dx \approx \frac{r_0}{2}\frac{B_1}{B_2}, \qquad dz \approx -\frac{r_0}{2}\frac{A_1}{B_2} \qquad \tan(2\theta) = -\frac{A_2}{B_2},$$

$B_n$, $A_n$, being the $n^{th}$ real and imaginary harmonics of the magnetic field $\vec{B}$. Note that these coefficients depend on the geometry of the coil. Here $r_0$ is a constant and $\theta$ the magnetic tilt.

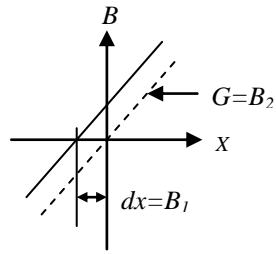

**Fig. 88:** Principle of zero detection applied to the *x* direction

Some imperfections due to the realization of the coil and of its support can occur: the geometry may not be exactly the one expected (Fig. 89). The lack of precision on the realization of the coil is not of first importance in the zero detection because of the symmetry of revolution of its influence. It should induce a scale factor to be applied to the residual $dx$ and $dz$ still existing after having set the definitive shims. Since these latter are small, that effect can be neglected on the alignment of the magnet.

The rotation axis of the coil may be outside the plane including the coil itself (Fig. 90). In that case, the flux seen by the coil is not strictly radial and a tilt error $d\theta = \delta/r$ appears on the orientation of the Qpole magnetic definition. Note that the case of an offset of the coil in the Z direction of about $\delta$ is strictly equivalent. That error can be estimated by rotating a magnet around its vertical axis and by using the average of the two corresponding tilt values.

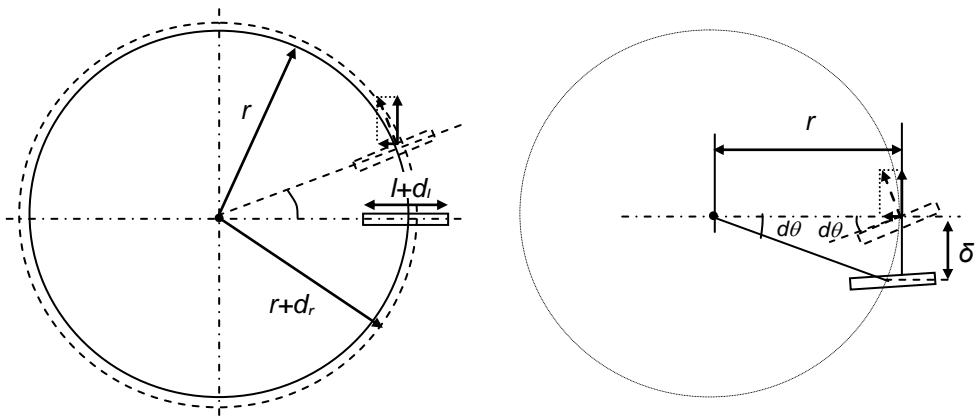

**Figs. 89 and 90:** Radial coil and geometrical defects

In the proposed case study, we shall suppose that the true axis of rotation is fixed and confused with the geometrical axis of the coil. In other words, there is no radial run-out due to the bearings.

Figure 91 shows the bench in detail: components and metrology loop (red dashed line). The rotation of the sensor is managed with an angular encoder. Its zero is defined by an inclinometer fixed on the sensor. It will be possible to link the angular readings to gravity and thus, measure the magnetic tilt of the magnet.

The magnetic sensor is fragile and a change during the period of two or three months when measuring all the Qpoles must be envisaged. Any other part of the bench could slightly change in position too, especially the pin which defines the X reference (10). As a conclusion, it is necessary to link the magnetic axis of the coil with respect to the X pin and to the girder surface which is the Z reference. A reference tool (3) has been designed: a permanent Qpole magnet with eight mechanical faces to position it on a stand fixed to the bench and in contact with the X pin and the Z surface (Fig. 91).

The eight positions allow a good redundancy: the measurements could describe a perfect octagon and their average is used as a final result. In addition, if the size of the reference tool is accurately known, it would be possible to use another tool in case of destruction of the first one. An ESL configuration appears here. The magnetic axis of the tool is defined with respect to its eight faces and to the coil. The use of the reference tool must be analysed in terms of STC: it allows avoiding $STC = (\mu m, \infty)$ for the whole metrology loop which is very difficult to achieve (no change possible during the campaign). If both tool and bench mechanics are supposed to be perfectly machined at the nominal sizes, then the result should be zero. It would be possible only if every part was machined at a high level of accuracy (better than 10 µm), that is very demanding work.

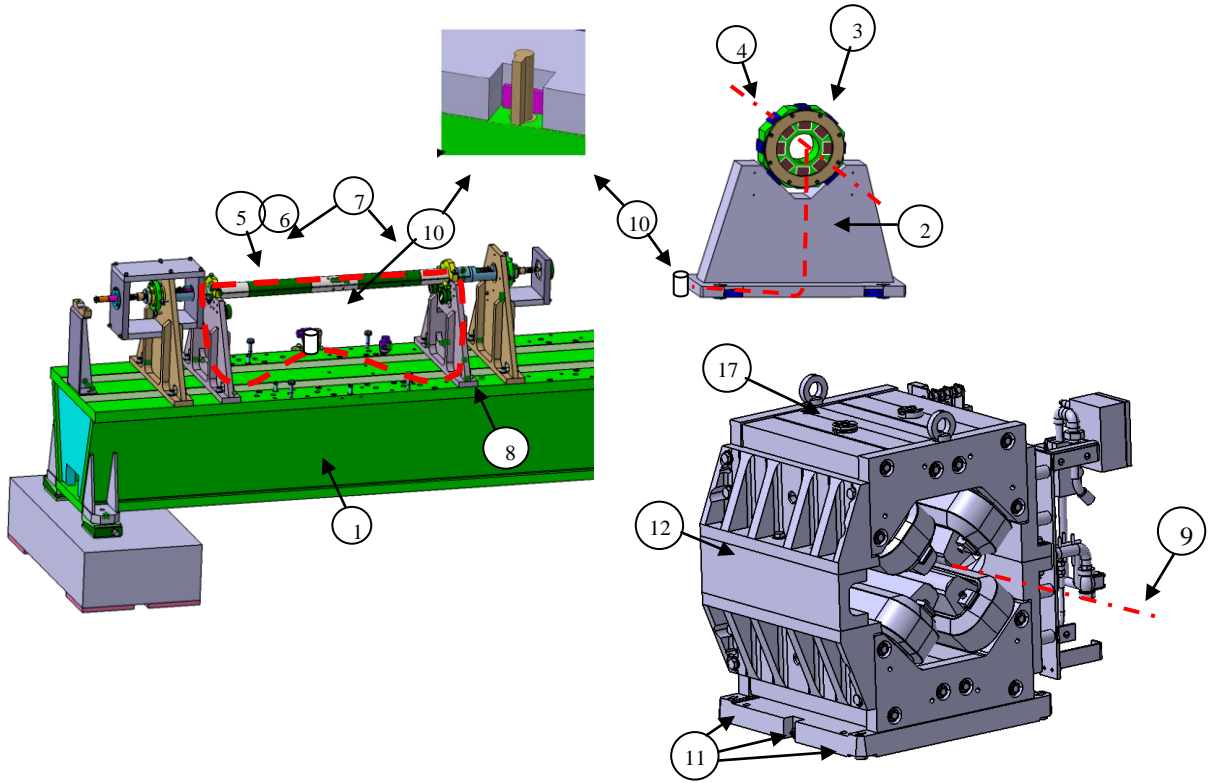

**Fig. 91:** Quadrupole bench for magnetic measurements: details and metrology loop (red dashed line)

The bench STC can be extrapolated by using its periodic survey with the same reference Qpole shown in Fig. 92 [23]. The magnet comparator has the following STC (according to the DOF), including its mechanical and electronic (sensors) stabilities (Table 1).

**Table 1:** Magnet comparator STC

| **X** | (few µm, 5 min) | rotation of the system |
|---|---|---|
| **Z** | (few µm, 3 months) | no rotation possible |
| **θs** | (10 µrad, 5min) | rotation of the system |

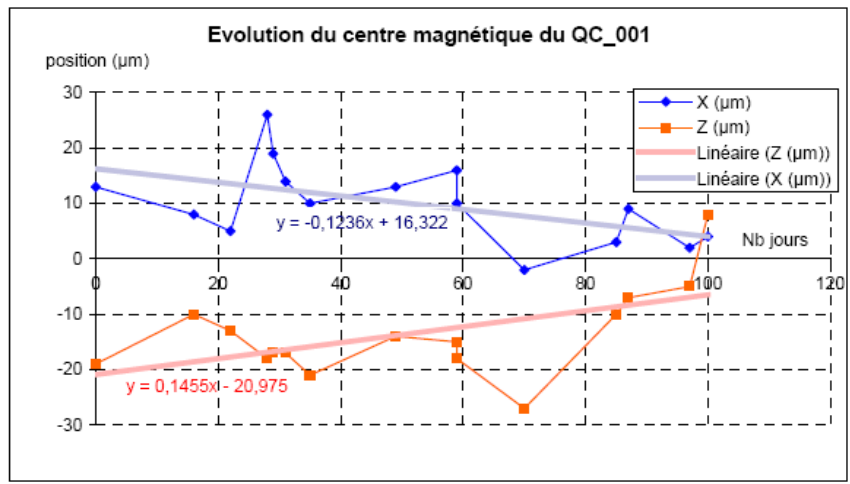

**Fig. 92:** Periodic magnetic checking of the BMS with a reference Qpole

**Table 2:** STC analysis of the bench

| No. | Component | Action | σ (µm) | Bias | STC = (µm,t) | STC = Easy/Diff. |
|---|---|---|---|---|---|---|
| 1 | **Bench** | *Mech.*\* | | | (,∞) | E |
| | | | | | | |
| 2 | **Tool stand** | | | | | |
| | | *Contact*\*\* | | ESL | (,min) | E |
| 3 | **Tool ext. face** | | | | | |
| | | *Mech.* | | | (,∞), (,min) = ESL | D,E |
| 4 | **Tool mag. axis** | | | | | |
| | | *Meas.* | 5 | | | |
| 5 | **Coil mag. axis** | | | | | |
| | | *Mech.* | | | (,∞) | E |
| 6 | **Coil rotat. axis** | | | | | |
| | | *Rotat.*\*\*\* | 5 | | (,∞) | E |
| 7 | **Ball bearings** | | | | | |
| | | *Contact* | | | (,∞) | E |
| 8 | **Coil stand** | | | | | |
| | | *Mech.* | | | (,∞) | E |
| 1 | **Bench** | | | | | |

\**Mech*: mechanical machining   \*\*\**Rotat*: Movement in rotation
\*\**Contact*: mechanical contact

### 10.3  Detection of the Qpole axis

Each Qpole is laid on the bench in order to be measured by the sensor. The whole set of magnets should 'see' the sensor and then they can be compared even if bench offsets still exist. The differential layout eliminates them. The shims are changed until the sensor measures a null field as it is on the magnetic axis of the magnet. The corresponding shim is the final one and will equip the Qpole.

The STC analysis must be done for the duration to measure all the magnets, i.e., equivalent to infinity. It shows that two weak points exist: the sensor and the radial pin. The first point is solved by the use of the reference tool. The second one comes from a lack of mechanical design: the pin has to withstand contact with magnets, whose weight is as much as 500 kg, approximately 200 times. Reaching $STC = (\mu m, \infty)$ is then very difficult. It is important to reliably link it to the bench.

The use of an inclinometer requires the definition of a referential for both STC analysis and calculation. In other words, the bench itself must not vary in inclination over a period of 2 or 3 months. For that, a second inclinometer must be fixed on the girder of the bench. All the inclinometry measurements will be normalized with respect to the slope variations of the bench.

## 10.4 Fiducialization

The fiducialization is materialized by a kinematical centring system (dot-line-plane) fixed on the upper face of the magnet. The set of functions 'dot–line' defines the *X*, *Z*, *tilt* references. The plane function helps with the centring of instruments and is defined by the yoke itself. The survey of the fiducials with respect to the magnetic definition of the Qpole is obtained on the magnetic bench by means of a stainless-steel structure called the 'Qpole comparator' (QC) (Fig. 93). This structure presents four electronic dial gauges, a pair at each end, one set in the vertical direction, the other one in the horizontal direction. In addition, an inclinometer is fixed on it. It allows the measurement of the magnetic tilt coming from the bench measurements through the Z shims of the magnet and the control of the lever arm of 365 mm inducing errors in the *X* direction of the fiducialization results.

**Table 3:** STC analysis of the Qpole axis detection

| No. | Component | Action | σ (µm) | Bias | STC = (µm,t) | STC = Easy/Diff. |
|---|---|---|---|---|---|---|
| 9 | **Qpole mag. axis** | | | | | |
| | | *Meas.* | 5 | | | |
| 5 | **Coil mag. axis** | | | | | |
| | | *Rotat.* | 5 | | (,∞) | E |
| 7 | **Ball bearings** | | | | | |
| | | *Contact* | | | (,∞) | E |
| 8 | **Coil stand** | | | | (,*Tool1\**) | |
| | | *Mech.* | | | (,∞) | E |
| 1 | **Bench** | | | | | |
| | | *Mech.* | 5 | | (,∞) | D |
| 10 | **Bench pin (surface)** | | | | | |
| | | *Contact* | | | (,∞) | E |
| 11 | **Qpole shim(s)** | | | | | |
| | | *Mech.* | 5 | | (,∞) | E |
| 12 | **Qpole yokes** | | | | | |
| | | *Mech.* | | | (,∞) | E |
| 9 | **Qpole mag. axis** | | | | | |

*Tool1* covers the items 5, 7, 8, 1, 10: Period between two uses of the Reference tool.

When the QC is centred on the magnet fiducials, the four dial gauges can touch the cylindrical support of the magnetic sensor. The sensor rotates a full revolution and the average of the 64 readings per dial gauge are stored, plus the inclinometer reading. The QC is then rotated by half of the revolution around its vertical axis, realizing a change of the horizontal dial gauges direction and of the inclinometer zero (note that the direction of the vertical dial gauges is not inversed). The measurements are once again carried out after permutation.

There is currently an ESL configuration: 2 × 64 points on the coil, 4 dial gauges and 1 inclinometer. These measurements are processed by using the average of the data between the two ends of the QC: it is actually, the magnetic centre of the magnet we want to define and not its real axis. The final result is a set of three offsets per Qpole: *dx, dz* and *dtilt*.

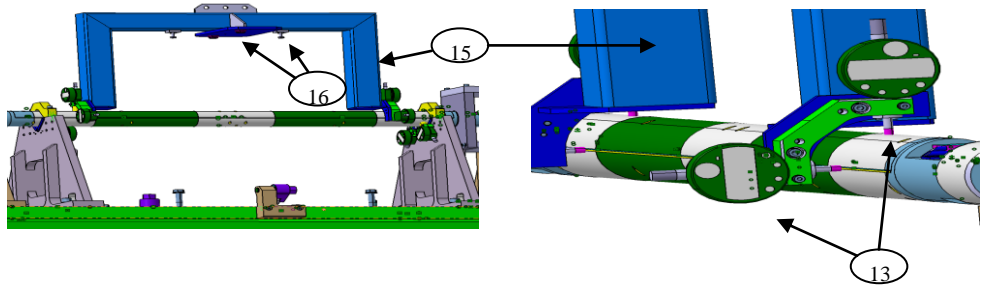

**Fig. 93:** Qpole comparator

The STC analysis raises the following sensitive points:

i) The *Z* measurements are not vertically permuted. As a consequence it is necessary to have a tool in order to check the vertical stability of the QC. A small and simple bench has been designed. It simulates the several contacts of the QC: centering system and dial gauges on an extremely rigid false Qpole. By using it, we can check the *Z* measurement of the QC including the electronic zero of the dial gauges at any moment. Practically speaking, this checking would be rather cumbersome on a regular basis. Since the stability of the structure is not so bad, we prefer to check only its link with the dial gauges before each use of the QC. The zero is far more unstable. A special wedge is dedicated to that operation.

ii) The QC is centred on the magnet by means of truncated spheres in contact with magnet fiducials. The distance between the two spheres must not change during the measurement campaign, i.e., the STC is ($\mu m$, $\infty$). A *dl* variation of that distance induces a *dl*/2 error on the results in the *X* direction.

iii) The bench tilt has to be measured each time the QC is used.

**Table 4:** STC analysis of fiducialization

| No. | Component | Action | σ (µm) | Bias | STC = (µm,t) | STC = Easy/Diff. |
|---|---|---|---|---|---|---|
| 9 | **Qpole mag. axis** | | | | | |
| | | *Meas.* | 5 | | (,30 min) | E |
| 5 | **Coil mag. axis** | | | | | |
| | | *Rotat.* | 5 | ESL | | |
| 13 | **Dial gauge** | | | | | |
| | | *Tool2*(14)* | 5 | | (,30 min) | E |
| 15 | **QC structure** | | | | | |
| | | X: *Mech.* | | | (,30 min) | E |
| | | Z: *Tool3**_ | | | (,days) | E |
| 16 | **Trunc. spheres** | | | | | |
| | | *Contact* | 5 | | (,30 min) | E |
| 17 | **Qpole fiducials** | | | | | |
| | | *Mech.* | | | (,∞) | E |
| 12 | **Qpole yoke** | | | | | |
| | | *Mech.* | | | (,∞) | E |
| 9 | **Qpole mag. axis** | | | | | |

*Tool2: Wedge
**Tool3: Z bench

## 10.5 Laser ecartometry of Qpoles mounted on a girder

The Qpoles are mechanically aligned by the contact of their shims with the girder: its upper surface is accurately machined for an altimetric reference (*Z*, *tilt*) and pins define the *X* position of the magnets.

That operation is a checking of the previous steps: by using the fiducials and their offsets measured with the QC, the Qpoles set should show a perfect alignment. It means that the metrology loop including the shimming operation, the fiducialization, and the laser survey with STR500 should be close to zero (Fig. 95).

Thus the result of this survey gives a good estimate of the quality of the work. The achieved results at SOLEIL were respectively in *X* and *Z*: 15 µm and 11 µm. A laser source is set at an extremity of the girder (Fig. 94). The retro-reflector sends back the beam to the telescope and its difference compared to the beam reference gives the *dx* and *dz* of the retro-reflector with respect to the beam. An inclinometer completes the survey to manage the lever arms. It is used twice with a rotation of 180°.

Since the measurements are carried out twice, at both ends for the laser, similar to a reversal, there is an ESL configuration for the laser ecartometer measurements.

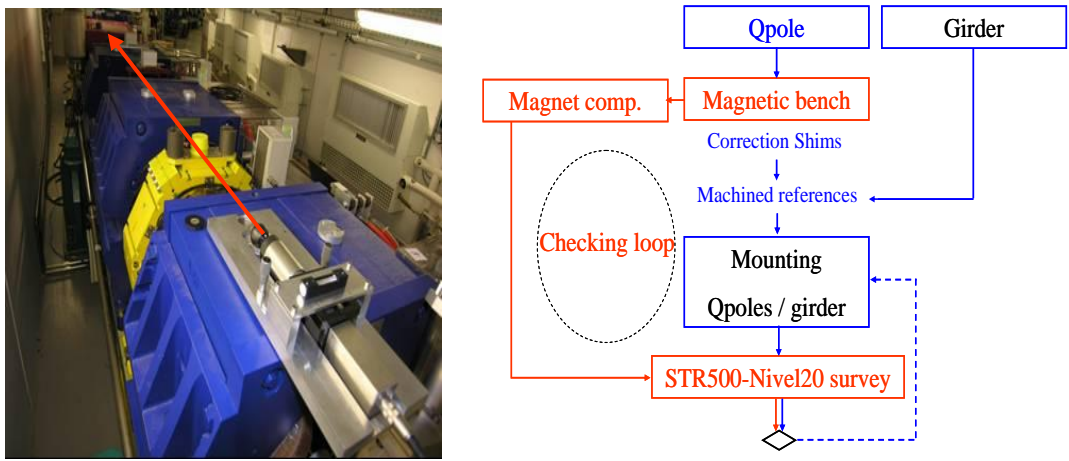

**Figs. 94 and 95:** Laser alignment and checking loop

**Table 5:** STC analysis of laser ecartometry

| No. | Component | Action | σ (µm) | Bias | STC = (µm,t) | STC = Easy/Diff. |
|---|---|---|---|---|---|---|
| 18 | **Laser beam** | ⇔* | ⇔ | ⇔ | (,2 min) ⇔ | ⇔ |
|  |  | *Meas.* | 5 | ESL |  |  |
| 19 | **Retro-reflector** |  |  |  |  |  |
|  |  | *Contact* |  |  | (,∞) | E |
| 17 | **Qpole fiducials** |  |  |  |  |  |
|  |  | *Fiduc.*** | 10 |  | (,∞) | E |
| 9 | **Qpole mag. axis** |  |  |  |  |  |
|  |  | *Bench**** | 10 |  | (,∞) | E |
| 11 | **Qpole shim** |  |  |  |  |  |
|  |  | *Contact* | 5 |  | (,∞) | E |
| 20 | **Girder pin (surface)** |  |  |  |  |  |
|  |  | *Mech.* | 5 |  | (,∞) | E |
| 21 | **Girder** | ⇔ | ⇔ | ⇔ | (,∞)⇔ | ⇔ |

*⇔      : Common to all magnets on a girder
**Fiduc.   : Fiducialization => offsets
***Bench   : Detection of the Qpole axis => Shim

## 10.6 Planimetric Qpole alignment (orbit definition)

The alignment of the Qpoles in the tunnel is carried out by iterations and leads to the final step, the accurate alignment with wire ecartometer and inclinometer. The proposed method is based on differential measurements to eliminate offsets error from the instrument. For more details, see Section 9.1, first example. The measures are common to two girders; it is a way to link them together in terms of a metrology loop.

A redundancy does not appear in the next table: all the girders are measured twice, with their two neighbours. The final least-squares calculation includes measurements from the laser ecartometry and also the TDA5005 measurements (accurate angles + distances). Since the orbit involves all the Qpoles, the metrology loop includes the girder stands and the concrete slab of the tunnel. Note that keeping $STC = (\mu m, \infty)$ for them is not possible, then a new survey and alignment operation whose frequency is called 'SA' in Table 6, will be necessary and depends on the level of the acceptable misalignment by the machine physicists.

**Table 6:** STC analysis of planimetric Qpole alignment

| No. | Component | Action | σ (µm) | Bias | STC = (µm,t) | STC = Easy/Diff. |
|---|---|---|---|---|---|---|
|  | **Wire** | ⇔ | ⇔ | ⇔ | (,10 min) ⇔ | ⇔ |
|  |  | *Meas.* | 10 |  |  |  |
| 22 | **Ecartometer centering** |  |  |  |  |  |
|  |  | *Contact* | 5 |  | (,10 min) | E |
| 17 | **Qpole fiducials** |  |  |  |  |  |
|  |  | *Fiduc.* | 10 |  | (,∞) | E |
| 9 | **Qpole mag. axis** |  |  |  |  |  |
|  |  | *Bench* | 10 |  | (,∞) | E |
| 11 | **Qpole shim** |  |  |  |  |  |
|  |  | *Contact* | 5 |  | (,∞) | E |
| 19 | **Girder pin** |  |  |  |  |  |
|  |  | *Contact* | 5 |  | (,∞) | E |
| 21 | **Girder** |  |  |  |  |  |
|  |  | *Mech.* |  |  | (50, SA)* | D |
| 22 | **Concrete slab** | ⇔ | ⇔ | ⇔ | (50, SA)* | D |

*SA: Period between two Survey and Alignment operations

## 10.7 Altimetric Qpole alignment (orbit definition)

Altimetry is obtained with HLS sensors whose reference is a free surface of water available all along the storage ring (Figs. 96 and 97). Linking the Qpole magnetic axis to that surface is a very sensitive operation. It requires the following steps in addition to the fiducialization:

– altimetric measurements from fiducials to HLS vessels,

– linking all the zero sensors together.

The first operation consists in measuring the Qpole fiducials of a girder and the three HLS vessels on the girder with the laser ecartometer and an inclinometer. A long rod laid on the vessels allows one to get the same approximate level for all the points. The laser ecartometer then measures only the vertical direction. The laser ecartometer is used twice, in two symmetrical positions related to the surveyed points. There is an ESL configuration for the laser ecartometer measurements. The STC requirement of the laser ecartometer is similar to the one for its use when checking the Qpole position on the girder.

The zero sensor link has already been described in Section 4.6, third example. Its frequency is once a year, the corresponding drift of sensor zeros is around 10 µm.

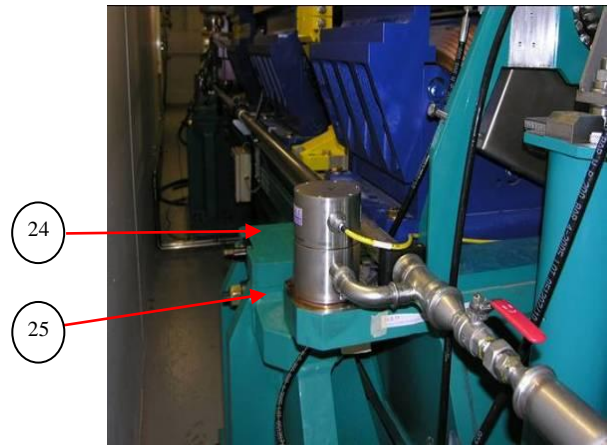

**Fig. 96:** HLS on girder

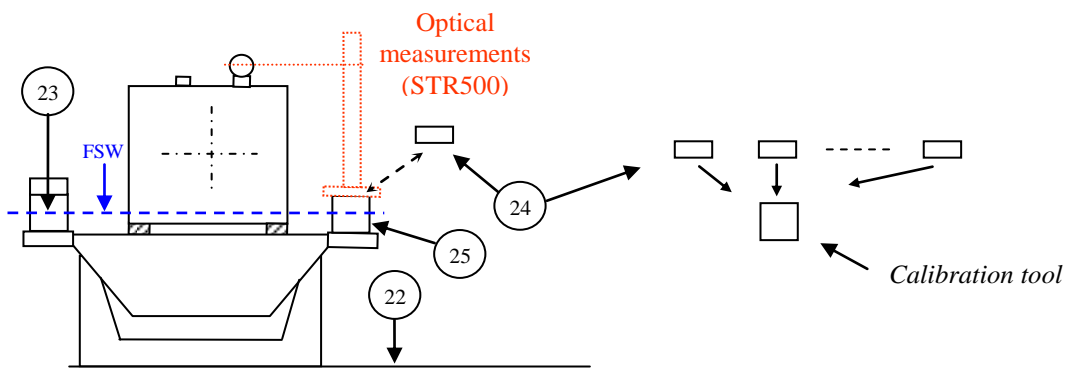

**Fig. 97:** HLS network and stainless-steel calibration tool

**Table 7:** STC analysis of altimetric Qpole alignment

| No. | Component | Action | σ (µm) | Bias | STC = (µm,t) | STC = Easy/Diff. |
|---|---|---|---|---|---|---|
| 22 | **Water** | ⇔ | ⇔ | ⇔ | ⇔ | ⇔ |
|  |  | *Meas.* | 5 |  | (,∞) | E |
| 24 | **HLS zero sensor** |  | 10 |  | (10,one year)* | E |
|  |  | *Contact* |  |  | (,∞) | E |
| 25 | **HLS vessel** |  |  |  |  |  |
|  |  | *Laser** | 10 |  | (,∞) | E |
| 17 | **Qpole fiducials** |  |  |  |  |  |
|  |  | *Bench* | 10 |  | (,∞) | E |
| 11 | **Qpole shims** |  |  |  |  |  |
|  |  | *Contact* | 5 |  | (,∞) | E |
| 22 | **Girder surface** |  |  |  |  |  |
|  |  | *Mech.* | 5 |  | (,∞) | E |
| 20 | **Girder** |  |  |  |  |  |
|  |  | *Mech.* |  |  | (50,SA)*** | D |
| 22 | **Concrete slab** | ⇔ | ⇔ | ⇔ | (50,SA) | D |

\* With the calibration tool.
\*\* Laser: Laser ecartometer measurements between HLS vessels and fiducials (not described here).
\*\*\* SA: Period between two survey and alignment operations

## 10.8 Planimetric alignment with precise tacheometer (orbit definition)

A tacheometer is an instrument based on a theodolite and equipped with an Electronic Distance Meter (EDM). It measures polar coordinates between two points. The Leica TDA5005 is used to measure the network of points defined by all the Qpole fiducials and wallbrackets. Each instrument for alignment operations has its own range of use. The TDA5005 operates at middle- and long-range of distances for pure alignment on a straight line and for the general shape of the storage ring. The latter cannot be detected with ecartometry.

The TDA5005 is centred on the fiducials of Qpole and brackets fixed on the SR walls. Qpoles are used in preference to benefit from angle accuracy. $STC_{\theta z} = (3.10^{-4}$ deg., 10 min). Every detail in the preparation of the campaign needs to reach only 10 min.

The redundancy of the measurements (Fig. 98) leads to a bundle adjustment based on least-squares criteria. Note that the metrology loop includes the slab and girders as the ecartometry and HLS measurements; the corresponding $STC = (50 \ \mu m$, SA), with SA being the duration between two realignment campaigns.

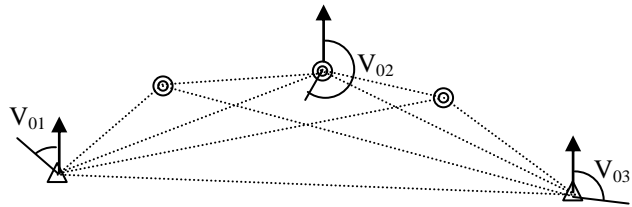

**Fig. 98:** Geodetic network

**Table 8:** STC analysis of planimetric survey

| No. | Component | Action | σ (µm) | Bias | STC = (µm,t) | STC = Easy/Diff. |
|---|---|---|---|---|---|---|
| 12 | **Qpole yoke$_{TDA}$** | | | | | |
| | | *Mech.* | | | | |
| 17 | **Qpole fiducials$_{TDA}$** | | | | | |
| | | *Contact* | 5 | | (,10 min) | E |
| 26 | **TDA centering** | | | | | |
| | | *Mech.* | | | | |
| 27 | **Zero (dist and angles)** | | | | (,10 min) | $D_\theta$** |
| 28 | **Air (dist and angles)** | *Meas.* | 0.12mm & $3.10^{-4}$deg | | | |
| 19 | **Retro-reflector** | | | | | |
| | | *Contact* | 5 | | (,10 min) | E |
| 17 | **Qpole fiducials/$_{refl}$** | | | | | |
| | | *Mech.* | | | | |
| 12 | **Qpole yoke$_{refl}$** | | | | | |
| | | *Mech.** | | | (50,SA)*** | D |
| 22 | **Slab** | | | | (50,SA) | D |
| | | *Mech.** | | | (50,SA) | D |
| 12 | **Qpole yoke$_{TDA}$** | | | | | |

\*: Including girders and stands
\*\*: Difficult for angle zero
\*\*\*SA: Period between two Survey and Alignment operations

**Acknowledgement**

I would like to thank Fabrice Marteau, member of the Magnetism and Insertion Group, SOLEIL, for his precious help on the topic of magnetic measurements.